\documentclass[nofootinbib,aps]{revtex4}

\usepackage{amssymb,amstext,amsmath}
\usepackage[dvips]{graphicx}
\usepackage{latexsym}
\usepackage{psfrag}
\usepackage{amsfonts}
\usepackage{bbm}
\usepackage{color}

\newcommand{\Ref}[1]{(\ref{#1})}

\newcommand{\eqa}{\begin{eqnarray}}
\newcommand{\neqa}{\end{eqnarray}}
\newcommand{\equ}{\begin{equation}}
\newcommand{\nequ}{\end{equation}}
\newcommand{\no}{\nonumber\\}

\newcommand{\N}{\mathbb{N}}

\newcommand{\hh}{{\cal H}}
\newcommand{\oo}{{\cal O}}

\def\om{\omega}

\def\la{\langle}
\def\ra{\rangle}
\newcommand{\bra}[1]{\la {#1}|}
\newcommand{\ket}[1]{|{#1}\ra}
\newcommand{\mean}[1]{\la{#1}\ra}
\newcommand{\6}{$\{6j\}$}

\def\d{\delta}
\def\f{\frac}

\newcommand{\p}{\partial}

\usepackage{bbm}

\newcommand{\scr}{\rm\scriptscriptstyle}

\newcommand{\SU}{\mathrm{SU}}
\newcommand{\SO}{\mathrm{SO}}
\newcommand{\Spin}{\mathrm{Spin}}

\newcommand{\lp}{\ell_{\rm P}}

\def\what{\widehat}
\def\aa{{\cal A}}
\def\ff{{\cal F}}
\def\hh{{\cal H}}
\def\mm{{\cal M}}
\def\oo{{\cal O}}
\def\pp{\partial}
\def\tt{{\cal T}}

\def\be{\begin{equation}}
\def\ee{\end{equation}}

\begin{document}

\title{\Large\bf Towards the graviton from spinfoams: \\
higher order corrections in the 3d toy model}
\author{{\bf Etera R. Livine}\footnote{elivine@perimeterinstitute.ca} }
\affiliation{Laboratoire de Physique, ENS Lyon, CNRS UMR 5672, 46 All\'ee d'Italie, 69364 Lyon,
France, and \\
Perimeter Institute, 31 Caroline St. N, Waterloo, ON N2L 2Y5, Canada.}
\author{{\bf Simone Speziale}\footnote{sspeziale@perimeterinstitute.ca}}
\affiliation{Perimeter Institute, 31 Caroline St. N, Waterloo, ON N2L 2Y5,
Canada.}
\author{{\bf Joshua L. Willis}\footnote{jwillis@mailaps.org}}
\affiliation{Department of Mathematics, The University of Western Ontario,
London, ON N6A 5B7, Canada.}

\begin{abstract}{
\noindent {We consider the recent calculation \cite{RovelliProp} of the graviton
propagator in the spinfoam formalism. Within the 3d toy model introduced in \cite{Io}, we
test how the spinfoam formalism can be used to construct the perturbative expansion of
graviton amplitudes. Although the 3d graviton is a pure gauge, one can choose to work in
a gauge where it is not zero and thus reproduce the structure of the 4d perturbative calculations.
We compute explicitly the next to leading and next to next to
leading orders, corresponding to one--loop and two--loop corrections. We show that while
the first arises entirely from the expansion of the Regge action around the flat
background, the latter receives contributions from the microscopic, non Regge--like,
quantum geometry. Surprisingly, this new contribution reduces the magnitude of
the next to next to leading order. It thus appears that the spinfoam formalism is likely
to substantially modify the conventional perturbative expansion at higher orders.

This result supports the interest in this approach. We then address a number of open
issues in the rest of the paper.  First, we discuss the boundary state ansatz, which is a key
ingredient in the whole construction. We propose a way to enhance the ansatz in order to
make the edge lengths and dihedral angles conjugate variables in a mathematically
well--defined way. Second, we show that the leading order is stable against
different choices of the face weights of the spinfoam model; the next to leading order,
on the other hand, is changed in a simple way, and we show that the topological face
weight minimizes it.  Finally, we extend the leading order result to the case of a regular,
but not equilateral, tetrahedron.}}
\end{abstract}

\maketitle


\section{Introduction}
The spinfoam formalism \cite{carlo} is a candidate covariant approach to a
non--perturbative quantisation of General Relativity (GR).  At present, it provides a
consistent background independent theory at the Planck scale, where it describes
spacetime as a discrete quantum geometry. However, the large scale behaviour of the
theory is less understood. Indeed, the formalism lacks a well--defined procedure to study
the semiclassical limit, to define particle scattering amplitudes and to reproduce
low--energy physics. In particular, considering the pure gravity case, we would expect to
recover in the low--energy regime the conventional perturbative expansion described in
term of gravitons. Recently, an important step towards this semiclassical limit has been
taken with a proposal for the construction of the 2-point function of quantum gravity
from the spinfoam amplitudes \cite{ModestoProp}. It relies on the use of the propagation
kernel (see for instance \cite{Mattei}) and on a boundary state peaked around flat
geometry. The proposal has been considered in 4d \cite{RovelliProp, noi} and 3d
\cite{Io} Riemannian GR without matter. In both cases the large scale limit of the
linearised 2-point function has been shown to reproduce the expected $1/p^2$ behaviour of
the free graviton, thus providing a first piece of evidence that spinfoams might
correctly lead back to general relativity in the semiclassical limit.


One of the appealing aspects of the proposal is that it can be used to compute not only
the free propagator, but also the quantum corrections due to the intrinsic non--linearity
of the theory. Such corrections have been studied in the conventional
(non--renormalisable) perturbative approach to quantum gravity, where they are understood
as graviton self--energies. In the (physically interesting) 4d case, these give rise to
corrections to the Newton potential (see for instance \cite{Donoghue,bohr}).
However, the perturbative approach can only be considered as an effective field theory, and not
as a complete theory, precisely because of its non--renormalisability:
an infinite counterterm arises at two loops \cite{Goroff}.
On the other hand, it has been argued that the cure to this non--renormalisabily could lie in the
use of non--perturbative approaches, such as the spinfoam formalism.
It is thus very interesting to compute the quantum corrections
to the free propagator within the spinfoam formalism,
both to give stronger support to the proposal of \cite{ModestoProp},
and to understand how full quantum gravity modifies the usual perturbative expansion of gravitons.

As a warm up exercise for the 4d case, we consider in this paper 3d quantum gravity and
study the emergence of higher order corrections to the free propagator. As is well
known, 3d general relativity (GR) has no local degrees of freedom. Due to this
peculiarity of the 3d case, the following clarification is necessary. If the
metric is quantised as a whole, the theory is exactly solvable \cite{Ponzano, Witten},
and this is the basis of the topological Ponzano--Regge (PR) model that we consider
here. However, if one treats the metric with the conventional perturbative expansion
$g_{\mu\nu}=\eta_{\mu\nu}+h_{\mu\nu}$, the field $h_{\mu\nu}$ is a pure gauge quantity:
consequently, the quantum theory does not have a propagating graviton \cite{Deser}. It is
nonetheless useful to consider the above expansion in a gauge where $h_{\mu\nu}$ is not
zero. Then the 2-point function,
$$
W_{\mu\nu\rho\sigma}(x,y)=\bra{0} \,{\rm T}
\left\{h_{\mu\nu}(x) h_{\rho\sigma}(y)\right\} \ket{0},
$$
can be evaluated in a perturbative expansion in $\lp$. The leading order, corresponding
to the free theory, goes as $1/p^2$ in momentum space, which in 3d means a $1/\ell$
dependence on the spacetime distance; analogously one can compute the self--energies, and
the perturbative expansion is not renormalisable \cite{Deser, Anselmi}, as in 4d. Therefore, it
makes sense to study in 3d how the spinfoam--defined 2-point function can cure the
non--renormalisability of GR by producing a different perturbative expansion than the
usual one.

To that end, we consider here a simple toy model introduced in \cite{Io}: the
contribution to the graviton propagator by a single tetrahedron belonging to a
discretisation of spacetime. A key advantage of this toy model is that it allows
efficient numerical simulations, which are substantially more
difficult in the 4d case. Indeed, it was the first numeric results of
\cite{Io} that gave hints of deviations at short scales from the
inverse power law $1/\ell$. There is a precise reason to expect that
such corrections would show a different structure than the
conventional perturbative expansion, and that is the modification of
microscopic structure by quantum geometry. To see why this can be
expected, recall first that since we are dealing with 3d Riemannian
GR, we consider the Ponzano--Regge spinfoam model. It is constructed
in terms of half--integers  (spins) labelling $\SU(2)$ irreducible
representations (irreps), variables which are related to geometric
quantities. The large scale limit then corresponds to the large 
spin limit. In that regime, the model is dominated by the path
integral for Regge calculus, with a specific choice of measure. Regge
calculus is a discrete approximation to classical GR, which captures
the non--linearity of the theory. Therefore, in this limit, one can
still expand around a flat background, and consider only small
perturbations around it. The leading order, which reproduces the
$1/\ell$ free graviton, is obtained by keeping only the quadratic term
in the Regge action, and the trivial background measure. Higher orders
in the action and in the measure give corrections with the structure
of the usual continuum perturbative expansion.  

Thus far, one would conclude
that the spinfoam formalism simply produces a somewhat discrete version of the
conventional theory. However, because the Regge path integral only emerges in the large
spin limit, the full theory predicts \emph{further} sources of corrections, arising from the
exact, non Regge--like, quantum geometry. Heuristically, these corrections are remnants
at large scales of the discrete microscopic structure of the spinfoam geometry. These
other corrections have no analogue in the conventional theory, and their presence is the
reason why we expect spinfoams to introduce new features in the perturbative expansion.
Indeed, it has often been suggested that the microscopic picture of quantum geometry
emerging from spinfoams would affect in some way the large scale dynamics, possibly
giving better finiteness properties. Here we show a clear example of how this could
happen. In particular, it is interesting to
study the interplay between the two different kinds of corrections. We
compute the next to leading (NLO) and the next to next to leading
(NNLO) orders, and we show that the quantum geometry corrections only
enter the NNLO, and moreover \emph{reduces its magnitude}. This
is the main result of this paper.

A key techical input to the expansion is the choice of boundary state entering
the definition of the 2-point function. In principle, the boundary state should be given
by the vacuum state of the theory, but in the absence of a well--defined prescription for it,
we must make an ansatz. In \cite{RovelliProp} a
Gaussian peaked around the (discrete quantities representing) the intrinsic and extrinsic
geometry was suggested for the boundary state, however there has been
little investigation of alternatives. So far, the only requirement on
the boundary state is that it be a good semiclassical state;
namely, that the relative uncertainties of the geometric quantities vanish in the large
spin limit. As the structure of the state is the same in both 3d and 4d, it is useful to
investigate its properties in the toy model here considered. The simple Gaussian
introduced in \cite{RovelliProp} has two problematic aspects: first, the variables
representing the intrinsic and the extrinsic geometry are not conjugate to each other in
a mathematically well defined way; second, it has an intricate $\mathbbm C$-structure,
due to the use of a complex phase. Here we show how both issues can be solved, by working
with a smooth function on $\SU(2)$, and using the harmonic analysis over the group. We
propose a new state and show that it approximates a Gaussian in the spin labels basis, and thus
satisfies the semiclassical requirements described above and leads to the same leading
order of the 2-point function. Enhancing the ansatz for the boundary state is our second
result, and we expect this to have implications especially in the 4d case. Furthermore,
we show that it is mainly the phase term that is responsible for the inverse power law
of the leading order of the 2-point function: the quadratic part of
the Gaussian serves mainly to damp the remaining oscillations of the PR kernel.

Another crucial point is the choice of the model; that is, the choice
of face and edge amplitudes. An important question is the stability of
results among different models. In fact, if one has a clear procedure to reproduce
low--energy physics from spinfoams, then that is a powerful tool to discriminate between
physically interesting models. To see how this can be done in practice, we consider a
particular class of models, which differ from the original PR model only in the weights
associated with the faces. Though restrictive, this generalisation is an interesting
example, as 4d models as well have some freedom in choosing the edge weights. Here we
show that changing the face weights does not change the leading order of the 2-point
function, but only the NLO. Furthermore, the NLO is changed in a very simple way, and it
is easy to show that whereas the original PR model (with topological face weight)
minimises the real part of the relative NLO correction, it is the trivial face weight
that minimises the absolute value of the NLO. This is our third result.

Finally, we also consider a geometric issue. Both in \cite{RovelliProp} and \cite{Io}, the boundary
geometry is represented by equilateral configurations, to simplify the geometric analysis.
That analysis becomes more involved for arbitrary configurations, but the
quantities to be computed are defined in the same way, and we expect the same results to hold.
To show that this hope is plausible, we consider a boundary geometry
of a regular---though not equilateral---tetrahedron.

The paper is organised as follows.
In the next section, we introduce the toy model and describe the physical setting. In section \ref{SectionNLO}
we study the perturbative expansion of the spinfoam 2-point function, and compute the leading order and NLO.
In section \ref{SectionNNLO} we compute the NNLO, and show that the spinfoam quantum geometry enters
at this order. In section \ref{sectionBoundary} we discuss the boundary state, and propose
a new ansatz. In section \ref{sectionk} we discuss the stability of the results obtained against
different face weights in the spinfoam model. In section \ref{sectionrett} we extend the construction to the case
of an isosceles tetrahedron.

Recall that in 3d the Newton constant $G$ has inverse mass dimensions (in units $c=1$);
we define the 3d Planck length as $\ell_{\rm P}=16\pi \hbar G$, and choose units $\lp=1$.

\section{Physical setting}
\label{sectionsetting}

The Ponzano--Regge spinfoam model allows us to compute the physical scalar product
between states of 2d geometry induced by 3d quantum gravity. This scalar product
implements the projection on the states satisfying the Hamiltonian constraint,
$\what{\hh}\,|\psi\ra=0$. The Hamiltonian constraint generates time
translations, and such states define the physical Hilbert space of the theory.
The standard setting is to consider a 3d manifold $\mm$ representing the
evolution in time of a 2d slice $\Sigma$ of constant topology, $\mm\,=\,\Sigma\times [0,1]$. We
choose two triangulations $\pp_i,\pp_f$ of the 2d slice $\Sigma$ on the initial and final
boundary and a triangulation $\Delta$ of $\mm$ which interpolates
between $\pp_i$ and $\pp_f$. In the following, we will assume for simplicity the initial
and final triangulations to be identical, $\pp_i=\pp_f=\pp$. A spin network state on
$\Sigma$, $|j_{e\in \pp}\ra$, is the assignment of the $\SU(2)$ representation labels
$j_e\in\N/2$ to the edges $e$ of the triangulation $\pp$. The kinematical scalar product
is naturally constructed such that two spin network states are orthogonal if their
$j_e$'s differ. On the other hand, the physical scalar product is defined as:
\be\label{scalar}
\la \psi_f|\psi_i\ra_{\rm ph}\,:=\,
\sum_{\{j_e\}}
\overline{\psi_f(j_{e\in\pp_f})}\,\aa_\Delta(j_e)\,\psi_i(j_{e\in\pp_i}),
\ee
where the sum is over all possible assignments of irreps to the interior edges of
the 3d triangulation $\Delta$, and the Ponzano-Regge amplitude
$\aa_\Delta(j_e)$ is defined as the product of Wigner's $\{6j\}$ symbols, each
associated to a tetrahedron $\tau$ of $\Delta$:
$$
\aa_\Delta(j_e)\,=\,
\prod_e (2j_e+1)\,
\prod_\tau \{6j\}_\tau.
$$
This state sum is topologically invariant and does not depend on the chosen triangulation
$\Delta$. In fact, the amplitude defined in this way usually gives an infinite result and
requires suitable gauge fixing \cite{diffeo}.

The physical observables of the theory are (gauge--invariant)
operators commuting with the Hamiltonian constraint, such as holonomy operators.
These observables are constants of motion, and we refer to them as complete observables,
in the terminology of \cite{RovelliPartial}. Using the scalar product \Ref{scalar},
one can compute the expectation value of a physical observable $\what\oo$,
$$
\la \psi_f|\what{\oo}\psi_i\ra_{\rm ph}\,=\,\la \what{\oo}\psi_f|\psi_i\ra_{\rm ph}.
$$

In the present work, we are interested in computing correlations between initial and
final states for the 2d metric. This corresponds in a sense to computing a
graviton propagator. Of course, the 2d metric is not gauge--invariant: these correlations
will be gauge--dependent and thus a priori unphysical. Nevertheless, we expect that the
insertion of particles in order to localize space points will allow to turn the 2d metric into
a complete observable and these correlations into physical quantities.

The two ingredients of our calculations are (i) working with partial observables measuring
the metric fluctuations and (ii) using a (partial) gauge fixing defining the proper time of
evolution between the initial and final slices. More precisely, the chosen partial observables
are geometric measurements. The simplest case to consider are the fluctuations of the
edge lengths (or of the angles) in the boundary triangulations:
the relevant observable is then simply the spin label $j_{e_0}$ of a given edge $e_0$,
or its deviation $\delta j_{e_0}= j_{e_0}-j_0$ with respect to some fixed reference
length $j_0$.
As we have stressed above, this correlation is gauge-dependent. To make it a physical
observable, two steps are required. First, we need to identify the two end points of the
edge under consideration with physical points, thus making the function a partial observable
\cite{RovelliPartial}. This is achieved by coupling gravity to a matter field: we insert
a particles at each of these two points (see \cite{PRI}) and express the boundary state $\psi$
in terms of the field propagator (see Section \ref{sectionBoundary} below).
The gauge fixing  amounts to fixing the representation labels of some
edges in the bulk $\Delta\setminus(\pp_i\cup\pp_f)$ in order to fix the (proper) time
between $\pp_i$ and $\pp_f$ to some value $T$. We choose a set $\ff$ of edges in the bulk
and set (by hand) the spins $j_e$ for edges $e\in\ff$ to some fixed value $J_e(T)$
depending on the value of the time $T$. The resulting PR amplitude can be written as
\equ
\aa^T_\Delta(j_e)\,=\,
\aa_\Delta(j_e)\,\prod_{e\in\ff}\delta_{j_e,\,J_e(T)}.
\nequ
This obviously breaks the topological invariance
of the spin foam amplitude. Nevertheless, as shown in
\cite{diffeo}, if $\ff$ is a tree $\tt$ (i.e. does not contain any loop), then this amounts
to a gauge fixing of the diffeomorphism invariance, which is at the origin of both the topological
invariance and the original divergence of the spinfoam amplitude.
We choose a 1-skeleton path along $\Delta$ from $\pp_i$ to
$\pp_f$ and fix the $j_e$'s along that path. That would correspond to fixing the proper
time of a point particle travelling along that path. More generally we could choose a
tree $\tt$ in the bulk and fix the $j_e$'s along that tree thinking once more of
particles travelling along each of these edges. Finally, in the generic case that $\Sigma$ is
open and has a boundary, we could insert particles on the boundary $\pp\Sigma\times[0,1]$
and fix their proper time and see how this propagates to the bulk.

Concretely, we choose a state $|\psi\ra$ and an observable $\what{\oo}$ with
which we define the correlation:
\be
\bra{\psi}\what{\oo}(0)\what{\oo}(T)\ket{\psi}\,\equiv\,
\f1Z{\sum_{\{j_e\}} (\what{\oo}\psi)(j_{e\in\pp_i})\,(\what{\oo}\psi)(j_{e\in\pp_f})\,\aa^T_\Delta(j_e)},
\qquad
{\rm with}
\quad
Z={\sum_{\{j_e\}} \psi(j_{e\in\pp_i})\,\psi(j_{e\in\pp_f})\,\aa^T_\Delta(j_e)}.
\ee
As discussed in \cite{noi}, $\ket{\psi}$ has to be a physical state of the boundary
geometry, namely a solution
of the full dynamics. Furthermore,
$|\psi\ra$ might be a coherent state, namely a state peaked around a configuration which
is a solution of the classical equations of motion. In the case of a free theory, for instance,
it could be some kind of Gaussian state.

We expect the two steps described above to
turn our correlation into a complete observable. This method of interpreting the boundary
as matter insertion will be investigated in future work \cite{inprep}; in particular, we will
see how it determines the 3d triangulation.

The hope is to
recover the classical values of these correlations in some limit. This would have two
important consequences. On the one hand, it would prove that we have the right semiclassical
behaviour, i.e that we recover gravity in a long distance regime.
On the other hand, it would allow
us to bridge between the discrete gauge--fixing that we use at the spinfoam level and
the gauge used to compute the correlations in the continuum case
(typically the temporal gauge plus the Coulomb gauge, see \cite{Io}). Clarifying these issues will definitely be very
useful when tackling the problem in 4d spinfoam gravity.

\medskip

\begin{figure}[ht]
\begin{center}\includegraphics[width=6cm]{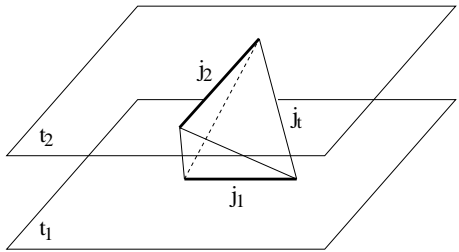}
\end{center}\caption{\small{The dynamical tetrahedron as evolution
between two hyperplanes, useful to study the correlation between edge lengths.
The labels give the physical lengths as $\ell_1=  C(j_1)$,
$\ell_2=  C(j_2)$, $\ell_t=  C(j_t)$, and $T=t_f-t_i=\ell_t/\sqrt 2$.}}\label{planes}
\end{figure}

In this paper, we will apply this framework to the simplest case. $\pp$ is made of a
single (open) edge and the 3d triangulation $\Delta$ is a single tetrahedron
interpolating between the initial edge and the final edge. Let us briefly recall the
setting of \cite{Io} (see also \cite{nutshell}). We consider a single tetrahedron
embedded in flat 3d Euclidean spacetime as drawn in Fig.\ref{planes}.
Using the PR model, the edge lengths
are given by $\ell_e= j_e+\f12 = C(j_e)$, where we take $C^2(j)=j(j+1)+\f14$ as the
Casimir of $\SU(2)$.
We assume to have measured the time $T=t_f-t_i$, and we are interested in computing the
correlation between
fluctuations, around a given background, of the length of the bottom edge and the length
of the top edge. This amounts to computing a single component of the graviton propagator,
such as $W_{1122}(T)$, choosing $j_1$ and $j_2$ to be along respectively the $x^1$ and $x^2$ axis.
The measured--time setting is obtained fixing the four bulk edges to some representation
label $j_t$, and it realizes a temporal gauge--fixing \cite{Io}.
The background around which we study the fluctuations is introduced by the state $\ket{\psi}$.
Assuming that $\ket{\psi}$ peaks $j_1$ and $j_2$ around a given value
$j_0$, we can compute the classical value $T$ in terms of $j_0$ and $j_t$. The generic case
is described in section \ref{sectionrett}. In the main body of the paper, we will restrict our analysis
to the equilateral tetrahedron for simplicity and take $j_t=j_0$. In that case, all edge
lengths are $\ell_0=C(j_0)$, and elementary geometry tells us that all dihedral
angles have the same value $\vartheta=\arccos(-\f13)$, and that $T=\ell_0/\sqrt{2}$.

\medskip

\begin{figure}[ht]
\begin{center}\includegraphics[width=7cm]{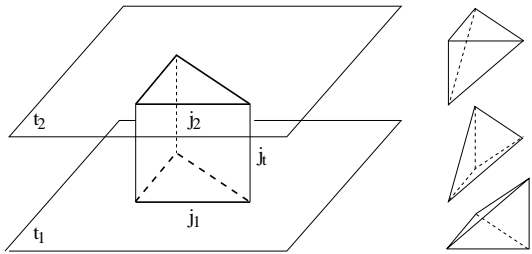}
\end{center}\caption{\small{The prism interpolating
between two hyperplanes, and its decomposition into 3 tetrahedra.
This setting can be used to study the correlation between triangle areas.}}\label{prism}
\end{figure}

The next case would be to consider correlations between the areas of two triangles. The 3d
triangulation would then be a prism made of 3 tetrahedra, as in Fig.\ref{prism}.
The temporal gauge--fixing is performed fixing the three ``vertical'' bulk edges, say to a value
$j_t$; this would
leave three ``diagonal'' edges with unspecified labels in the bulk, which we will have
to sum over to compute the correlations. The state $\ket{\psi}$ will introduce a background
value for the $j$'s in the bottom and top triangles, say two equilateral triangles with all
edge lengths $\ell_0$. We now have nine sums to perform, thus this setting
is computationally much more involved than
the single tetrahedron setting, and for this reason we leave it for future investigation. We point out
nevertheless that this would shed light on the following issues:
\begin{itemize}
\item By allowing to compute more correlations, this setting will give the full tensorial structure of the graviton propagator. We could
then compare it to the classical result and check the precise correspondence between the
discrete gauge fixing and the gauge fixing in the continuum.
\item It is the simplest setting in which we can study the effects of many tetrahedra
in the bulk.
\item We would be able to analyze how the bulk edges which we have not fixed to $j_t$ get
peaked (or not) around their classical values. This would show explicitly how fixing the
time on the boundary propagates into the bulk.
\end{itemize}
From now on we focus on the computation of the graviton propagator in the framework of the
single tetrahedron, studying the behaviour of its leading order and developing the necessary
tools to extract the quantum gravity corrections.

\section{The free propagator and the next to leading order correction}
\label{SectionNLO}

\subsection{Computation of the leading order}

We consider the single tetrahedron model described in the previous Section.
To study the perturbative expansion of the 2-point function in a Coulomb--like gauge--fixing,
our starting point is the following quantity,
\equ\label{Wexact}
W_{1122}(j_0) = \f1{j_0^4}\,\f1{\cal N}\, \sum_{j_1, j_2=0}^{2j_0} \left[\prod_{i=1}^2 \, (2j_i+1)
\;[C^2(j_i) - C^2(j_0)] \, \Psi_0[j_i] \right]\,
\left\{ \begin{array}{ccc} j_1 & j_0 & j_0 \\ j_2 & j_0 & j_0 \end{array} \right\}.
\nequ
The normalisation factor $\cal N$ is given by the same sum appearing above, without the
two Casimir insertions $[C^2(j_i) - C^2(j_0)]$. The latter represent the expectation
values of the metric perturbation $h_{\mu\nu}$ on the boundary spin network.
$\Psi_0[j_i]$ is the boundary state, representing the vacuum state of the theory (see the
discussion in \cite{noi}), and the \6 symbol is the PR kernel. For the motivations
leading to the expression above to be interpreted as the 2-point function, see \cite{Io}.
As for the boundary state, we assume for the computation of the free propagator the
following Gaussian ansatz:
\equ\label{psi0}
\Psi_0[j_i]=\exp\left\{-\f{\alpha}{2} \d j_i^2 + i \vartheta (j_i+\f12) \right\},
\nequ
where $\d j_i = j_i-j_0$, and $\alpha=\f4{3j_0}$ (see \cite{Io}). The essential point is
the $1/j_0$ dependence of $\alpha$ which leads to a Gaussian state with a width growing
in $\sqrt{j_0}$. Remarkably, this matches the scaling of the uncertainty in measuring lengths in 3d quantum gravity, $\delta j \sim \sqrt{j}$ or equivalently $\delta l \sim \sqrt{l \lp}$ \cite{Livine:2005fr}.
The exact coefficient $4/3$ leads to a precise physical interpretation
of the 2-point function as an harmonic oscillator, but changing it would not modify
the asymptotic behaviour of the 2-point function. We will discuss later in
section \ref{sectionBoundary} the physical meaning of this ansatz.

The large scale regime of \Ref{Wexact} can be studied using
the well known asymptotics of the \6 symbol \cite{Ponzano, asympt},
\equ\label{asymp}
\left\{ \begin{array}{ccc} j_1 & j_0 & j_0 \\ j_2 & j_0 & j_0 \end{array} \right\}
= \f{\cos\left( S_{\rm R}[j_e] +\f\pi 4  \right)}{\sqrt{12\, \pi\, V(j_1, j_2, j_0)}} +o({j^{-\f32}}),
\nequ
where $V(j_1,j_2,j_0)$ is the volume of the tetrahedron,
\equ
V(j_1, j_2, j_0) = \f1{12}\sqrt{4 \, \ell_0^2\, \ell_1^2\, \ell_2^2
- \ell_1^2 \, \ell_2^4 - \ell_1^4 \, \ell_2^2},
\nequ
with $\ell_i=j_i+\f12$, and it reduces to $V_0=\f{\ell_0^3}{6\sqrt2}$ in  the equilateral case.
$S_{\rm R}[j_e]$ is the Regge action,
\equ\label{regge}
S_{\rm R}[j_e] = 
\sum_{e}\, (j_e+\f12) \, \theta_e(j_e),
\nequ
where the $\theta_e$ are dihedral angles, defined as the angles between the external normals
to the triangles. In the Appendix, we give their expression in terms of edge lengths. The Regge
action is a discretised version of GR, which captures the non--linearity of the theory.
We can then expand it around flat spacetime, precisely as in the continuum. In our calculation,
this is induced by the Gaussian \Ref{psi0} entering \Ref{Wexact}; the expansion
around the background value $j_0$ for both $j_i$'s reads
\equ\label{Regge1}
S_{\rm R}[j_e] = S_{\rm R}[j_0] 
+ \vartheta  \,(\d j_1 + \d j_2)+
\f{1}{2} \sum_{i,k=1}^2 \f{\p^2 S_{\rm R}}{\p j_i \p j_k}{\Big|_{j_e=j_0}} \d j_i \d j_{k}
+\f16\sum_{i,k,l=1}^2 \f{\p^3 S_{\rm R}}{\p j_i \p j_k \p j_l}{\Big|_{j_e=j_0}} \d j_i \d j_{k} \d j_{l}+\ldots,
\nequ
where $S_{\rm R}[j_0] = 6\,\vartheta\, (j_0+\f12)$. We write this as
\equ
S_{\rm R}[j_e] = \sum_{a,b=0}^\infty \f1{a! b!} \, M_{ab} \, \d j_1^a \, \d j_2^b,
\nequ
with
$M_{ab}:=\f{\p S_{\rm R}}{\p j_1^a \p j_2^b}{\Big|_{j_e=j_0}}.$
This matrix can be evaluated from elementary geometry, and the elements which
are relevant to the following computations are reported in the Appendix.
Indeed, we have $M_{ab}\sim j_0^{1-a-b}$, thus in the $j_0\mapsto \infty$
limit higher derivatives of the Regge action become negligible, and this justifies
the expansion.

As noted in \cite{RovelliProp}, the key point here is that
the zeroth and first order terms in \Ref{Regge1}
reproduce $\vartheta(j_1+j_2+1)$, which is the phase of \Ref{Wexact} induced by
\Ref{psi0}.
Therefore, when we use the expansions \Ref{asymp} and \Ref{Regge1} in
the exact expression \Ref{Wexact}, the phase coming from \Ref{psi0}
is cancelled or doubled, depending on the sign
of the two exponentials of the cosine. But because the phase makes the argument of the sum
rapidly oscillating, we expect only the exponential where the phase is cancelled
to contribute to the sum.
This intuition is indeed confirmed by numerical simulations, and we can thus write
\equ\label{Wbis}
W_{1122}(j_0) = \f1{j_0^4}\,\f1{\cal N} \sum_{j_1, j_2}  \;
\f{\prod_i \, (2j_i+1)\;[C^2(j_i) - C^2(j_0)]\;\Psi_0[j_i] }{\sqrt{V(j_1,j_2,j_0)}}\,
\exp\left\{-iS_{\rm R} \right\},
\nequ
where we have absorbed the constant factor $\f{e^{-i\f\pi4}}{4\sqrt{3\pi}}$ in the normalisation.

Let us comment this first result. The exact expression for the 2-point function is given in
\Ref{Wexact}. By keeping only the first order in the large $j$
expansion of the \6 symbol, equation \Ref{asymp}, we are able to rewrite \Ref{Wexact}
as \Ref{Wbis}. This expression is what we would write down to compute
the 2-point function using not the PR kernel, but directly the path integral for Regge calculus,
choosing the factor
\equ\label{measure}
\mu(j_e):=\f{\prod_i \, (2j_i+1)}{\sqrt{V(j_1,j_2,j_0)}}
\nequ
for the measure of the path integral. Notice that this measure is largely determined by the volume.
The Regge path integral corresponds to a discretised version of quantum gravity,
and it has a highly non--linear structure.
If we now expand for large $j_0$ the measure and the Regge action, this corresponds to
a discrete analogue of the continuum perturbative expansion of quantum gravity.
The first non trivial contributions, corresponding to the free propagator, come from the quadratic
term in the expansion \Ref{Regge1} of the action, and from the trivial background term $\mu(j_0)$ in the measure.
Using the explicit value of $\vartheta$, one indeed finds \cite{Io} the following leading order,
\equ\label{LO}
W^{\scr LO}_{1122}(j_0) = \f32 \f{e^{i\vartheta}}{j_0} =-\f1{j_0} (\f12-i\sqrt2).
\nequ
This result can be interpreted as the contribution from a single tetrahedron to
the free propagator $\f{e^{i\om T}}{2\om}$, with $\om \simeq \f8{3j_0}$ (see \cite{Io}),
and it is supported by numerical calculations, reported in Fig.\ref{fit} below.

The scheme of the perturbative expansion leading to this
free propagator can be summarised in the following box:
\begin{center}
\framebox{\parbox{10cm}{
\begin{tabular}{ccccc}
\Ref{Wexact} & {\Ref{asymp}} & \Ref{Wbis} & \Ref{Regge1} & \Ref{LO}$+\ldots$
\\ quantum gravity & ${\longrightarrow}$ & Regge path integral &
${\longrightarrow}$ & linearised theory \\
\end{tabular}}}
\end{center}
From the scheme above, it should be clear that we have two sources of corrections to \Ref{LO}:

\renewcommand{\labelenumi}{(\roman{enumi})}
\begin{enumerate}
\item contributions coming from higher orders in the expansion of the Regge action and the
measure;

\item contributions coming from higher orders in the expansion of the \6 symbol.
\end{enumerate}

We will keep this notation (i) and (ii) throughout the rest of the paper. Corrections (i)
come from the non--linear structure of the Regge path integral; whereas the background
measure and the quadratic action provide the $1/j_0$ behaviour, non trivial contributions
from the measure and higher orders in the expansion
\Ref{Regge1} of the Regge action provide corrections in higher powers of $\lp$. Because
of their origin, these corrections can be considered as the  discrete analogues of the
QFT self--energies.

On the other hand, (ii) are new kind of corrections, arising from deviations of the \6
symbol from the asymptotics \Ref{asymp}; these corrections have no QFT analogues, and
cannot be guessed by discretising GR \emph{\`a la} Regge. In other words, the microscopic
discrete geometry, and its non--Regge like dynamics, does affect the large scale behaviour
in a non--trivial way. In particular, we show below that (ii) does not contribute to the
NLO, but arises only at the NNLO.

An important remark should be added here. The boundary state $\Psi_0[j]$ in \Ref{Wexact}
is in principle the vacuum state of the full theory, just as the \6 is the kernel of the full theory.
In computing the free propagator above, we assumed the Gaussian form \Ref{psi0}. However, when
computing the corrections to the free propagator, we expect contributions from $\Psi_0[j]$
beyond \Ref{psi0}. At the present state of investigation,
the exact form of the boundary state is not fully understood,
thus in the following we discard these terms.\footnote{Of course, we expect
these contributions to be non--negligible, and the reason we do not consider them here
lies in the lack of control we have over the vacuum state of the theory,
together with the exploratory spirit of this paper.} See also the comments at the end
of Section \ref{sectionBoundary}.

\subsection{NLO: contributions from the Regge action}

Let us begin by studying the corrections of type (i); namely, we start from the Regge
path integral \Ref{Wbis} and study the corrections arising from the measure and terms
higher than quadratic in \Ref{Regge1}. To make the following calculations clearer, let us
rewrite \Ref{Wbis} collecting the quadratic terms in the exponent,
\equ\label{W1}
W_{1122}(j_0) = \f1{j_0^4}\,\f1{\cal N} \sum_{j_1, j_2}
F(j_0, \d j_1, \d j_2)
\,\exp\left\{ -\f{1}{2} \sum A_{ik} \d j_i \d j_{k}\right\},
\nequ
where we have introduced the shorthand notation
\equ\label{F}
F(j_0, \d j_1, \d j_2):=
\mu(j_e) \prod_i \, [C^2(j_i) - C^2(j_0)]
\exp\left\{-i \sum_{a+b\geq 3}^\infty \f1{a! b!} \, M_{ab} \, \d j_1^a \, \d j_2^b \right\},
\nequ
and we have absorbed a constant factor $\exp\{-i 4\,\vartheta\, (j_0+\f12)\}$
in the normalisation. The normalisation is obtained by taking the same sum in \Ref{W1} and substituting
$F(j_0, \d j_1, \d j_2)$ with
\equ\label{FN}
F_{\cal N}(j_0, \d j_1, \d j_2):=
\mu(j_e) \exp\left\{-i \sum_{a+b\geq 3}^\infty \f1{a! b!} \, M_{ab} \, \d j_1^a \, \d j_2^b \right\}.
\nequ

Using the explicit values of $\alpha$ given above and of the second derivatives of the Regge
action reported in the Appendix, the quadratic exponent can be shown to be
\equ\label{A}
A_{ik} = \alpha \d_{ik}+i \f{\p^2 S_{\rm R}}{\p j_i \p j_k}{\Big|_{j_e=j_0}}
= \f4{3j_0} \left( \begin{array}{cc} 1+i \cot\vartheta & -\f i{\sin\vartheta} \\ \\
-\f i{\sin\vartheta} & 1+i \cot\vartheta \end{array}
\right).
\nequ
In the following calculations, we will need the inverse of this matrix, which is
\equ
A^{-1}_{ik} = \f{3j_0}8 \left( \begin{array}{cc} 1 & e^{i\vartheta} \\ \\
e^{i\vartheta} & 1 \end{array}
\right),
\nequ
where $e^{i\vartheta}=-\f12+i\sqrt{2}$. All of our
analytic calculations will keep track of this simple algebraic form.

The fact that the dominant term in the exponent is quadratic suggests that
we can evaluate the sums \Ref{W1} by approximating them with Gaussian integrals.
In fact, in the large $j_0$ regime, we have
$
\sum_{j_i=0}^{2j_0}\sim \int_0^{2j_0} d j_i = \int_{-j_0}^{j_0} d \d j_i.
$
As is well known,
\equ
\int_{-\infty}^{\infty} d x \, e^{-\f1{2j_0}x^2} -
\int_{-j_0}^{j_0} d x \, e^{-\f1{2j_0}x^2} \sim e^{-\sqrt{j_0}},
\nequ
thus approximating the sums with Gaussian integrals does not affect the NLO of \Ref{W1}.
At this point, the logic to study the NLO order in $j_0$ is clear: we expand \Ref{F}
and \Ref{FN} in powers
of $\d j_1$ and $\d j_2$, and then we evaluate the integrals as momenta of Gaussians.
If we expand \Ref{F}, we get
\eqa\label{Fexpanded}
F(j_0, \d j_1, \d j_2) &=& \f1{\sqrt V_0}
\left[(2j_0+1)^4 \d j_1 \d j_2 +
(2j_0+1)^3 \f{10 j_0+11}{4 (j_0+1)} (\d j_1 \d j_2^2+\d j_1^2 \d j_2)+ \right.\no
&& \left.+ (2j_0+1)^2\f{(84j_0^2+124j_0+57)}{32 (j_0+1)^2}
(\d j_1 \d j_2^3+\d j_1^3 \d j_2) + (2j_0+1)^2 \f{116j_0^2+236j_0+125}{16 (j_0+1)^2}
\d j_1^2 \d j_2^2 +\ldots\right] \times \no &&
\left[1-i \sum_{a+b=3}^\infty \f1{a! b!} \, M_{ab} \, \d j_1^a \, \d j_2^b
-\f12 \left(\sum_{a+b=3}^\infty \f1{a! b!} \, M_{ab} \, \d j_1^a \, \d j_2^b\right)^2 +\ldots\right].
\neqa
The structure of this expansion is rather intricate, because the coefficients go down as
$j_0\mapsto\infty$, whereas the integrals (approximating the sums) go
up. In more detail, we have that on the one hand, the coefficients of
$\d j_i^k$ in the expansion of 
the measure go as $\f1{\sqrt{V_0}}j_0^{6-k}$, and $M_{ab}\sim j_0^{1-a-b}$; on the other
hand, because the squared width of the Gaussian, and thus the values of the squared
momenta, are proportional to $j_0$, each factor $\d j_i^2$ in the integrand contributes a
factor $j_0$ after the integration is performed. By inspecting \Ref{Fexpanded}, we see
that the leading term is of the form $j_0^5/\sqrt{V_0}$, and it comes from the first term
in both the expansion of the measure and the integral. The next to leading is of the form
$j_0^4/\sqrt{V_0}$, and it has a number of contributions: terms up to the fourth order in
the expansion of the measure, together with the first, second (with $a+b=3,4$) and third
(with $a+b=3$) terms in the expansion of the exponential contribute. Namely, terms up to
the fourth order in the expansion of both the measure and the Regge action enter the NLO.

The same analysis for the expansion of the  normalisation $\cal N$ shows that we need
only the second order in the expansion of the measure, and still the fourth in the
expansion of the action,
\eqa\label{FNexpanded}
F_{\cal N}(j_0, \d j_1, \d j_2) &\simeq& \f1{\sqrt V_0}
\left[(2j_0+1)^2 -\f12(2j_0+1)(\d j_1+\d j_2) +
\f{17}8 (\d j_1^2 + \d j_2^2)+ \f54 \d j_1 \d j_2 \right]\times \no &&
\left[1-i \sum_{a+b=3, 4} \f1{a! b!} \, M_{ab} \, \d j_1^a \, \d j_2^b
-\f12 \left(\sum_{a+b=3} \f1{a! b!} \, M_{ab} \, \d j_1^a \, \d j_2^b\right)^2 \right].
\neqa

At this point, we evaluate separately the Gaussian integrals in the numerator and in the
denominator. We then collect all terms in decreasing powers of $j_0$, thus obtaining
the following structure,
\equ\label{exp}
W_{1122}(j_0)=\f1{j_0^4}\f{b_1 j_0^5+ b_2 j_0^4 + \ldots}{B_1 j_0^2+B_2 j_0 + \ldots}
= \f{1}{B_1}\left[\f{b_1}{j_0}+\f1{j_0^2}(b_2-\f{B_2}{B_1}b_1)+\ldots\right].
\nequ
In particular, $b_1=\f{3e^{i\vartheta}}{2\sqrt{\det A}}$, $B_1=\f1{\sqrt{\det A}}$, and
$\f{b_1}{B_1}$ gives  the LO \Ref{LO} discussed in the previous Section. The
NLO can be now computed to be
\equ\label{NLOn}
W_{1122}^{\scr NLO(i)}=-\f1{j_0^2}\f{7481 - i 1048\sqrt{2}}{5184 }=
-\f1{j_0^2}( 1.4431 - i 0.2859),
\nequ
where we have again used $\vartheta=\arccos(-\f13)$. The main point to
remember is that this is the contribution to the NLO coming from
corrections of type (i). 

\begin{figure}[ht]
\begin{center}\includegraphics[width=6cm,angle=-90]{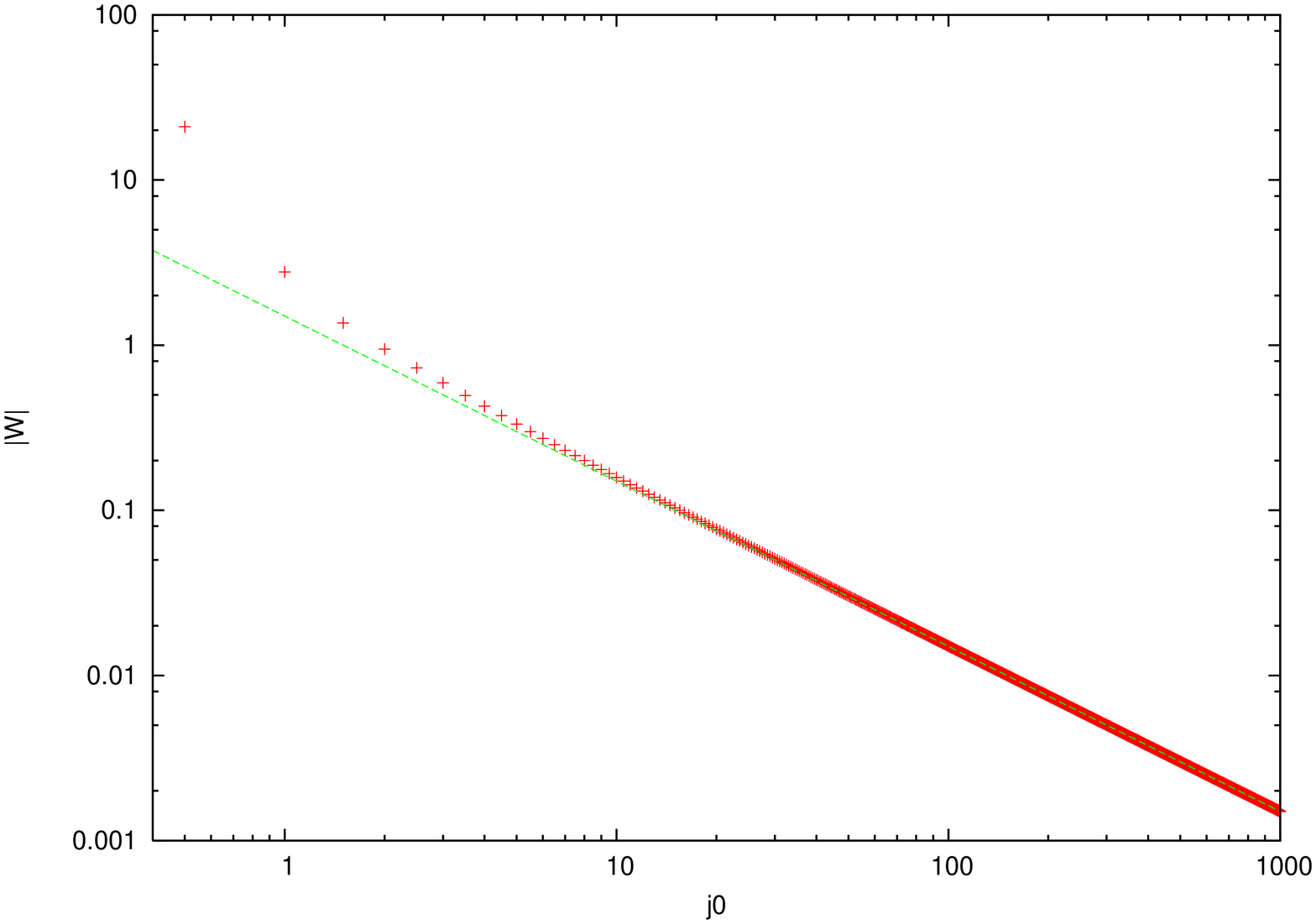}
\includegraphics[width=6cm,angle=-90]{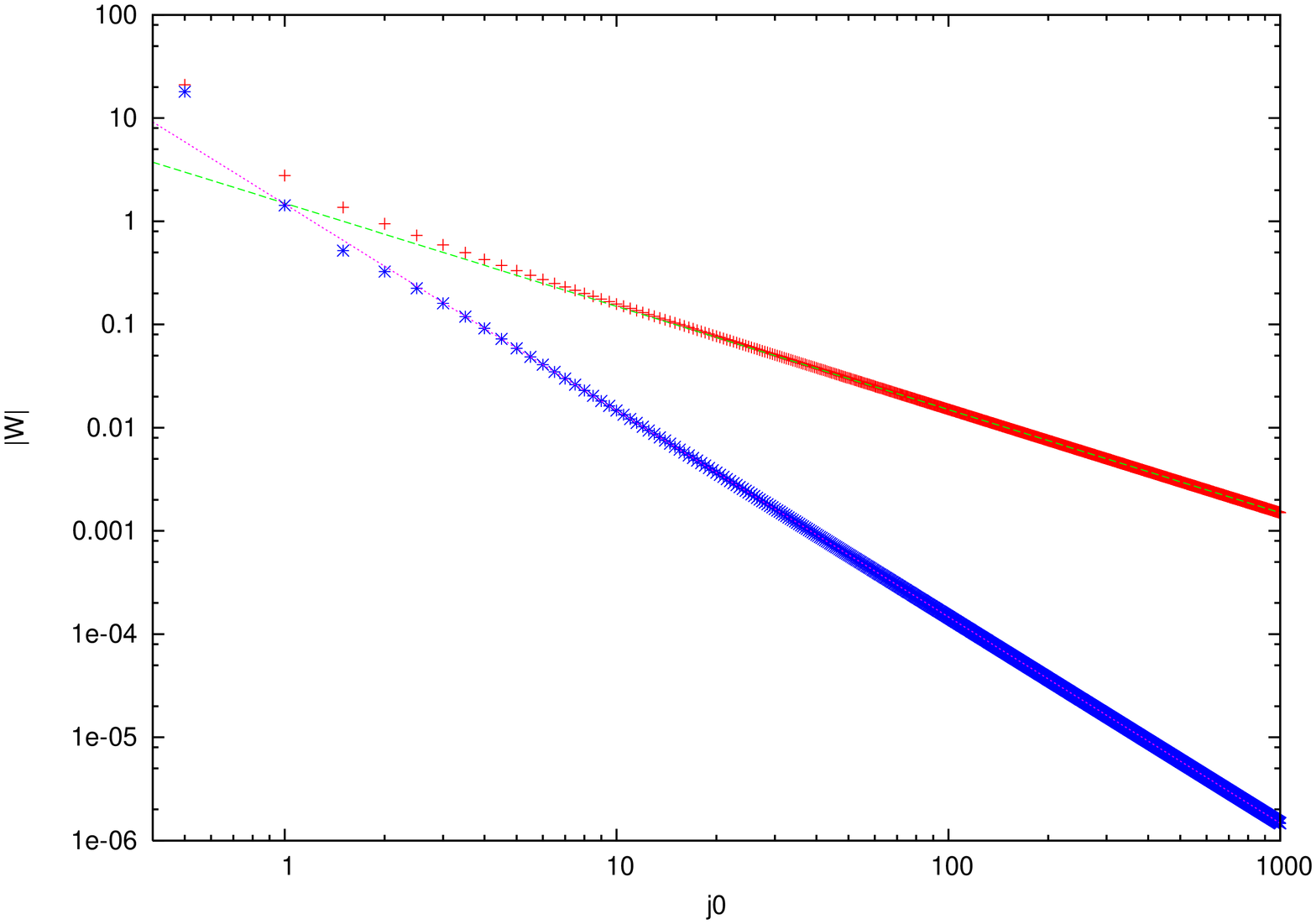}
\end{center}\caption{\small{The plots showing the numerical evaluation of the 2-point function.
Left plot: the leading order. In red, the numerical evaluations of
 \Ref{Wexact}; the dashed green line is the analytic result \Ref{LO}. Right plot: the leading order is plotted again,
 this time together with the next to leading order. The numerical evaluations of the latter is in blue,
 whereas the analytic result \Ref{NLOn} is the red dashed line.}}\label{fit}
\end{figure}

To support this analytic calculation, we fit the NLO using the numerical evaluation of
the exact formula \Ref{Wexact}, and subtracting the fit for the LO \Ref{LO}. The result
of the fit is a behaviour $1/{j_0^2}$ with numerical coefficient ${-1.4431+i 0.2860}$
(see Fig.\ref{fit}). This matches our analytic calculation up to three digits.
Therefore, there is room for corrections of type (ii) at NLO only if these are very
small, of magnitude $10^{-4}$.

\subsection{NLO: contributions from the \6}
We now consider (ii), the contribution coming from higher orders in the
asymptotics of the \6 symbol \Ref{asymp}. From the considerations at the end of the
previous Section, we know that if these corrections are present, they are indeed
very small. We begin the investigation by  studying
numerically the next term in the asymptotics of the \6 symbol for the equilateral
case, when all spins are $j_0$ and $S_{\rm R}[j_0] = 6 (j_0+\f12)\vartheta$.
We use the following ansatz,
\equ\label{asymp2}
\left\{ \begin{array}{ccc} j_0 & j_0 & j_0 \\ j_0 & j_0 & j_0 \end{array} \right\}
=  \f{1}{2^{1/4}\sqrt{\pi} \, j_0^{3/2}} \cos\left( S_{\rm R}[j_0] +\f\pi 4  \right) +
\f{a_2}{j_0^{5/2}}\cos\left( S_{\rm R}[j_0] +\varphi  \right) +o({j_0^{-\f52}}),
\nequ
and we perform a two parameters fit, finding the values
$a_2=-0.3575$, $\varphi=0.6843$. The good performance of this fit suggests the simple ansatz
\equ\label{ansatz}
\f{a_2}{(6\sqrt2)^{\f56}} \f{\cos\left( S_{\rm R}[j_1, j_2, j_0] +\varphi  \right)}{V(j_1,j_2,j_0)^{\f56}}
\nequ
as the second term in the expansion \Ref{asymp}. If we assume this ansatz,
we see that the second term in the asymptotics goes
as $j_0^{-5/2}$. This is just one power below the first term, thus in principle we should expect
the second term of the asymptotics to contribute to the NLO of $W_{1122}(j_0)$, which is
precisely one power below the LO. Notice also that $a_2$ is of the same magnitude as
$\f{1}{2^{1/4}\sqrt{\pi}}\sim 0.4744$,
so that if the second term does contribute to the NLO, it will do so with the same magnitude
as the contribution (i). However, we know that there is no room for this, as the value computed
from (i) already matches the exact numerical value up to $10^{-4}$.

We conclude that corrections coming from the second term do not enter the NLO, i.e.
\equ
W_{1122}^{\scr NLO}\equiv W_{1122}^{\scr NLO(i)}.
\nequ
Understanding the exact way this happens would require a more complete study of the
second term in \Ref{asymp2}, which is beyond the scope of this
paper.\footnote{To give a feeling for how this might happen, we note
the following. Using the ansatz \Ref{ansatz}, which is the simplest
homogeneous extrapolation of the second term in \Ref{asymp2}, the expansion around the equilateral background
would be
\equ\label{asymp3}
\f{a_2}{(6\sqrt2)^{\f56}} \f{\cos\left( S_{\rm R}[j_1, j_2, j_0] +\varphi  \right)}{V(j_1,j_2,j_0)^{\f56}}
= \f{a'}{j_0}\f{\cos\left( S_{\rm R}[j_0] +\varphi  \right)}{V_0^{\f32}}+\ldots
\nequ
where $a'=\sqrt{(6\sqrt2)}\,{a_2}$.
The expansion \Ref{exp} would now look like
\eqa\label{ciccio}
W_{1122}(j_0)&=&\f1{j_0^4}\f{b_1 j_0^5+ b_2 j_0^4 + \ldots+
\f{a'}{j_0}\left[b_1' j_0^5+ b_2' j_0^4 + \ldots\right]}
{B_1 j_0^2+B_2 j_0 + \ldots +\f{a'}{j_0}\left[B_1' j_0^2+B_2' j_0 + \ldots\right]}
= \f{1}{B_1}\left[\f{b_1}{j_0}+\f1{j_0^2}\Big(b_2+a'b_1'-(B_2+a'B_1')\f{b_1}{B_1}\Big)+\ldots\right].
\neqa
It is easy to see that \Ref{asymp3} does not contribute at NLO, but only at NNLO.
In fact, the effect of this new term simply amounts
to a change in the measure \Ref{measure}, thus
$b_1'=b_1$ and $B_1'=B_1$; therefore the NLO of \Ref{ciccio} is $\f1{B_1}(b_2-\f{B_2}{B_1}b_1)$, unchanged
by the addition of \Ref{asymp3}.
On the other hand, $b_2'\neq b_2$ and $B_2'\neq B_2$, because these terms are sensible to different measures;
therefore \Ref{asymp3} contributes to the NNLO. We will see in the next section that this is indeed the case.}

\section{NNLO: the pure quantum gravity effects}\label{SectionNNLO}

To study the NNLO, we proceed as in the previous section.
We first compute analytically the corrections of type (i) coming from
expanding the measure and the action, and we then compare the result with the
numerical fit obtained from \Ref{Wexact}, to see if corrections of type (ii) enter.
The analytic calculations can be performed as above, thus we do not
report here the somewhat tedious details. However, the complexity of the calculations
grows as the number of contributions increases largely with every order;
we notice that for the NNLO, in the expansion \Ref{Fexpanded},
we need up to the eighth derivative in the expansion of the measure,
and up to the sixth derivative in the Regge action (namely $a+b\leq 6$); similarly,
we need up to the sixth derivative in both the measure and the action in
the expansion of $F_{\cal N}$.
The result  we obtain is
\equ\label{NNLOn}
W_{1122}^{\scr NNLO(i)}=\f1{ j_0^3}\f{62758+i 127789\sqrt{2}}{248832}\simeq
\f1{ j_0^3} (0.2522 + i 0.7263).
\nequ
As expected, the NNLO is one power below the NLO. Notice
that the magnitude of the real part is one order of magnitude smaller than the NLO, and the magnitude
of the imaginary part same order. This is going to be changed by (ii), the corrections
coming from the quantum gravity kernel.

As before, we now evaluate numerically the NNLO, by subtracting \Ref{LO} and \Ref{NLOn} to \Ref{Wexact}.
The best fit gives
\equ
W_{1122}^{\scr NNLO(tot)} =W_{1122}^{\scr NNLO(i)}+W_{1122}^{\scr NNLO(ii)}
\simeq \f1{ j_0^3} (-0.0225+i 0.0730).
\nequ
This fit is quite different from \Ref{NNLOn}, which means that the corrections (ii) do enter the NNLO.
Notice that, as expected from \Ref{asymp2}, (i) and (ii) are
of the same magnitude. What is remarkable, is that (ii) have the neat effect of \emph{reducing
the magnitude of the correction}.
At this point it becomes interesting to evaluate analytically the second order in  \Ref{asymp2},
and then compute exactly the NNLO coming from it, $W_{1122}^{\scr NNLO(ii)}$. We leave this issue open for further
development.

Let us now use this result to draw some conclusions. Our analysis of the NLO has pointed
out that it arises only from higher order corrections in \Ref{Wbis}. Namely, the NLO is a
result of the Regge path integral alone, and is not sensitive to
deviations arising from the spinfoam quantum geometry. The latter, on
the other hand, \emph{are} important at NNLO, where
they contribute with the same order of magnitude as the Regge--like
corrections, but with opposite sign,
so that the total NNLO coefficient is smaller than that of the NLO. The structure of the corrections is
summarized as follows:

\begin{center}
\begin{picture}(100,110)(0,-100)
\put(-90,0){\framebox{full propagator}}\put(-45,-10){$\vector(0,-1){20}$}
\put(-80,-85){\framebox{\parbox{3.5cm}{{\hspace*{0.1cm}quadratic term in the \\ \hspace*{0.3cm}Regge path integral}}}}
\put(-20,-65){$\vector(0,1){20}$}
\put(0,0){\framebox{Regge path integral corrections}}\put(15,-10){$\vector(0,-1){20}$}
\put(40,-90){\framebox{\parbox{6.5cm}{\center{Regge path integral corrections \\ and \\
spinfoam quantum geometry corrections}}}}\put(50,-65){$\vector(0,1){20}$}
\put(-50,-40){$W = W^{\scr LO} + W^{\scr NLO} + W^{\scr NNLO} + \ldots$}
\end{picture}
\end{center}

\medskip

It would be very interesting if a similar structure emerges in 4d as well.
To this regard, we conclude this section with a speculative remark.
It has been argued that background independent approaches could be the key to
avoid the non--renormalisability of GR. The PR model considered here is one
such approach, and 3d GR, if treated perturbatively by expanding $g_{\mu\nu}=
\eta_{\mu\nu}+h_{\mu\nu}$, suffers the same infinities as the 4d case
(in a gauge where $h_{\mu\nu\neq 0}$).
Thus, it makes sense to address this idea within the 3d case.
In a discretised picture, the conventional perturbative framework is equivalent to using only
the Regge path integral, namely starting our computations from \Ref{Wbis}.
So, in a sense, the perturbative expansion that includes only corrections
of type (i) is the discrete equivalent of the conventional non--renormalisable
perturbative expansion. However, because we did not start from \Ref{Wbis},
but from the exact background independent quantity \Ref{Wexact}, we find that new
corrections will come in at higher orders, thus changing the structure of
the perturbative expansion. Could this provide a possible mechanism for renormalisability?
It is very speculative to say so, however we find it intriguing that the
new corrections enter at NNLO, precisely where the  non--renormalisability of GR shows
up.\footnote{In \cite{etera} it has been shown that the spinfoam quantum geometry
indeed affects the propagation of point particles. Interestingly, it does so in such a way
that the new dynamics can be encoded in new Feynman rules. It would be interesting to study
if the modifications produced here can be as well reabsorbed in modified Feynman rules.}

\section{On the boundary state}\label{sectionBoundary}

As should be clear from the calculations above, the boundary state plays a key
role for the extraction of the inverse power law behaviour, thus it is important to
motivate the choice
\Ref{psi0}. In principle, the boundary state should be given by the vacuum state of
the full quantum theory; however, such a state is not known in LQG. To overcome this lack
of knowledge, one can proceed with an ansatz. To compute the free propagator, in
\cite{RovelliProp} it was proposed to take, as the vacuum state for the linearised theory,
a Gaussian state peaked around the values of the labels for both the intrinsic and the
extrinsic geometry of a chosen background, by analogy with quantum mechanics. This
suggestion is realized in 3d by \Ref{psi0}. Indeed, we can think of it as a state for a
quantum mechanical system of conjugate variables $j$ and $\phi$, representing discretized
intrinsic and extrinsic geometry. Then, the main property satisfied by \Ref{psi0} is to
peak the values of those variable around, respectively, $j_0$ and $\vartheta$, and to
minimize, in the limit $j_0\mapsto\infty$, the relative uncertainties $\f{\mean{\Delta
j}}{\mean{j}}$ and $\f{\mean{\Delta \phi}}{\mean\phi}$. Notice that in order for this last
property to be true, it is crucial the $1/j_0$ dependence of the inverse width $\alpha$.

Below we show how this construction can be made more precise, by looking at
$j$ and $\phi$ as conjugate variables as defined by the Fourier transform
on the structure group of the PR model. Before doing that, we discuss some
interesting aspects of the phase term in \Ref{psi0}.

\subsection{The phase term}

\begin{figure}[ht]
\begin{center}\includegraphics[width=6cm,angle=-90]{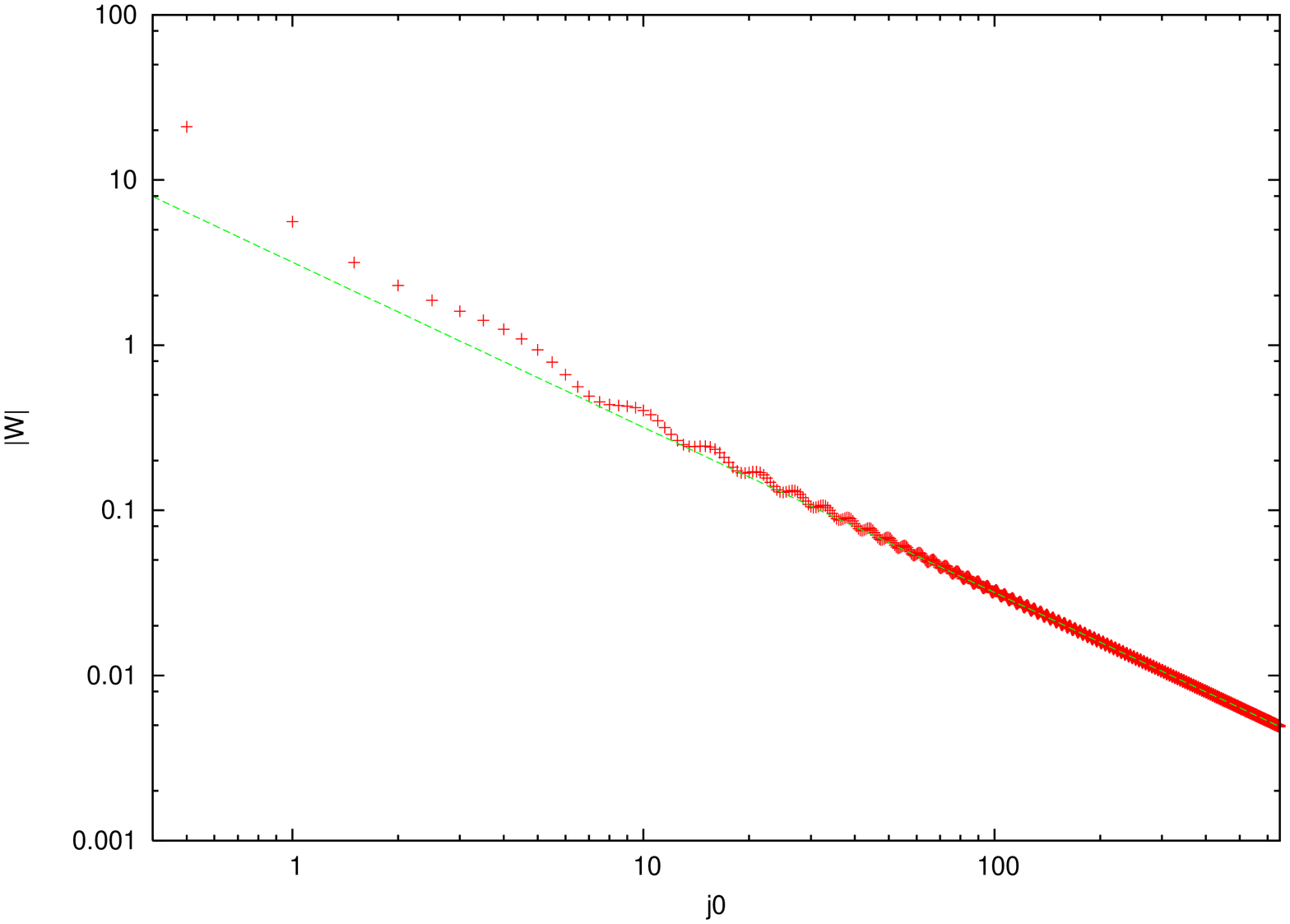}
\includegraphics[width=6cm,angle=-90]{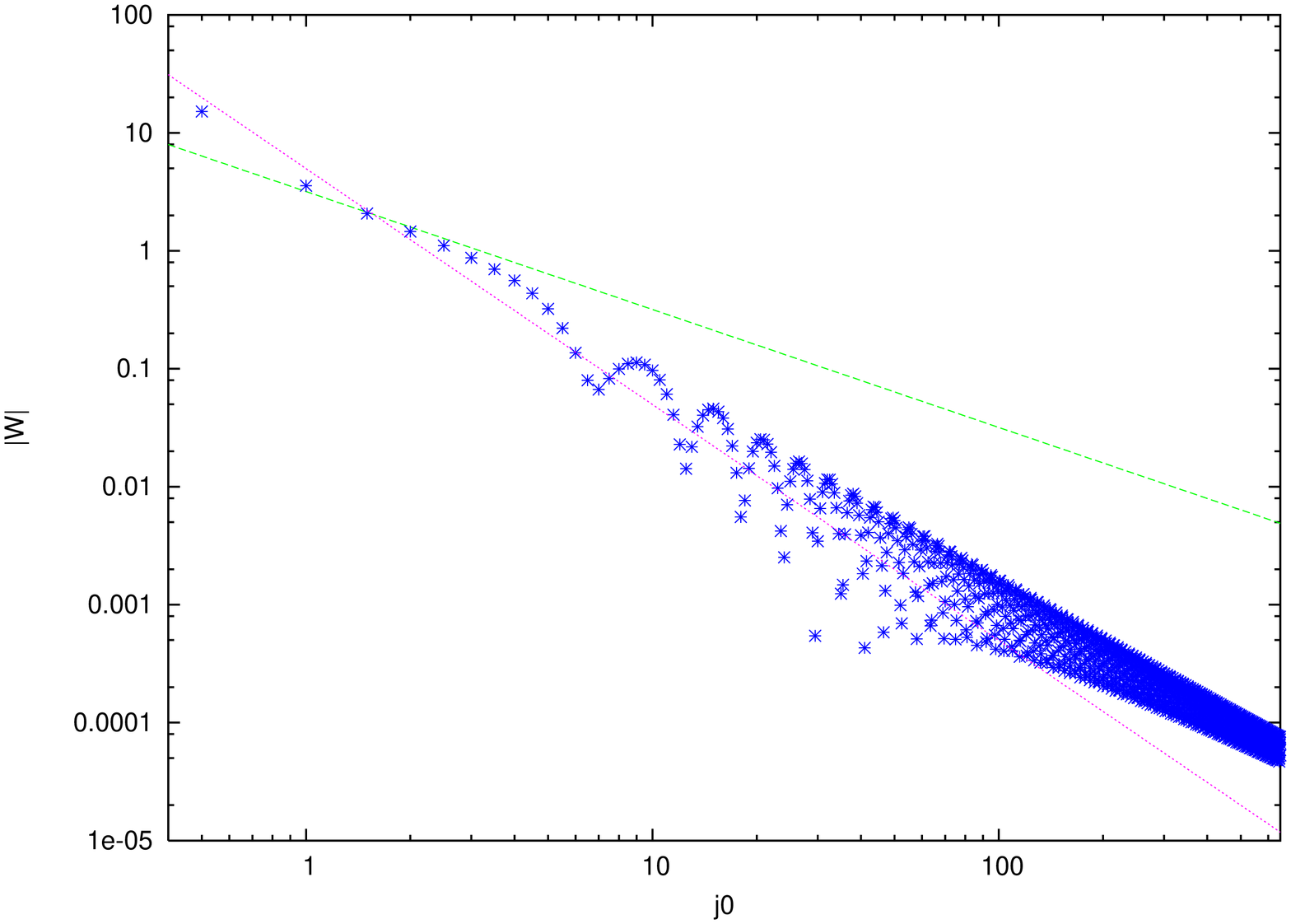}
\end{center}\caption{\small{The plots showing the numerical evaluation of the 2-point function
with $\alpha=0$. Left plot: the leading order. In red, the numerical evaluations of
 \Ref{Wexact}; the dashed green line is the analytic result in
 \Ref{zero}. Right plot: the numerical evaluations of the next to leading order, in blue,
 together with the first two terms of the analytic result \Ref{NLOn}. The green dashed line
 is the leading order, whereas the red dashed line shows the next to leading.}}\label{fitalpha0}
\end{figure}

We start our investigation by noting an important fact: it is the phase term of \Ref{psi0},
\equ\label{fase}
\Phi_\vartheta(j)=e^{i(j+\f12)\vartheta},
\nequ
that is mainly responsible for the $1/j_0$ behaviour of the asymptotics. To understand
this, recall that the phase term of \Ref{psi0} represents a state whose dihedral angles
are peaked around the equilateral configuration. Therefore, the presence of the phase
term is enough to expand the Regge action around such a configuration; but since the bulk
edges of the tetrahedron are fixed to $j_0$, the equilateral configuration is exactly the
one used in the calculations in Section \ref{SectionNLO}. This means that even if we set
$\alpha=0$ in \Ref{psi0}, we can proceed consistently with the perturbative expansion as
in Section \ref{SectionNLO}. The only difference is that now the matrix $A$ in \Ref{A}
consists only of the second derivatives of the Regge action. Indeed, we find the
following analytic result:
\equ\label{zero}
W_{1122}(j_0) \simeq \f1{j_0}i\f9{2\sqrt{2}}+\f1{j_0^2}\Big(-\f{245}{64}+i\f9{2\sqrt{2}}\Big).
\nequ
Once again, we support this result with the numerical simulations, which are reported in Fig.\ref{fitalpha0}.
Notice that we get much more oscillation, due to the absence of the
quadratic damping factor.
In particular, we see from the right figure that the NLO oscillates
quite strongly, and this makes it hard to fit its value.
Of course, \Ref{psi0}
with $\alpha=0$ loses the semiclassical properties described above, however, this is
just an exercise to show that actually is the phase term the key object, as much as
the calculations are concerned.

The fact that the phase term alone is sufficient to get the right
leading--order asymptotic behaviour,
suggests that we have freedom in choosing the angle: with $\alpha=0$, different angles
would still give non trivial results. Indeed, these would correspond to non equilateral
geometric configurations (see Section \ref{sectionrett} below).

The phase term has another interesting property. By inspecting \Ref{Wexact}, we see that
it is the only $\mathbbm C$-valued quantity entering the definition of the 2-point
function; thus, it is responsible for the 2-point function being a complex quantity.
Recall that we are working in a measured--time setting, where the 2-point function is
reconstructed as $\f{e^{i\om T}}{2\om}$, so a complex result is what we need in order to
be consistent with this interpretation. However, one can also consider a different
setting, where all boundary edges are allowed to fluctuate. This  ``general boundary''
context is particularly relevant in the 4d case model discussed in \cite{RovelliProp},
where the measured--time setting is not applicable.\footnote{This can be easily
understood, as the setting described in section \ref{sectionsetting} (see in particular
Fig.\ref{planes}) does not generalize to 4d: there is no embedding of a 4-simplex in 4d
Euclidean space such that two triangles of the 4-simplex lie in different parallel
hyperplanes.} In the general boundary context, one expects to reconstruct the 2-point
function in configuration space; in the Riemannian case, a real function of the spacetime
distance between the points, so that the computation of the 2-point function has to lead
to a real result. This is not obvious, if the phase term is $\mathbbm C$-valued. In fact,
notice that in \cite{RovelliProp}, the author has to choose a $\mathbbm C$-valued
(non--diagonal) quadratic term in the Gaussian
 boundary state, in order to compensate for the complex phase term, and make the final result real.
Here, we would like to suggest that the reality of the result can be obtained keeping the
simple quadratic Gaussian term, but rather changing the phase to a real quantity. We
argue below that this can be done without losing the $1/j_0$ behaviour.

\begin{figure}[ht]
\begin{center}\includegraphics[width=6cm,angle=-90]{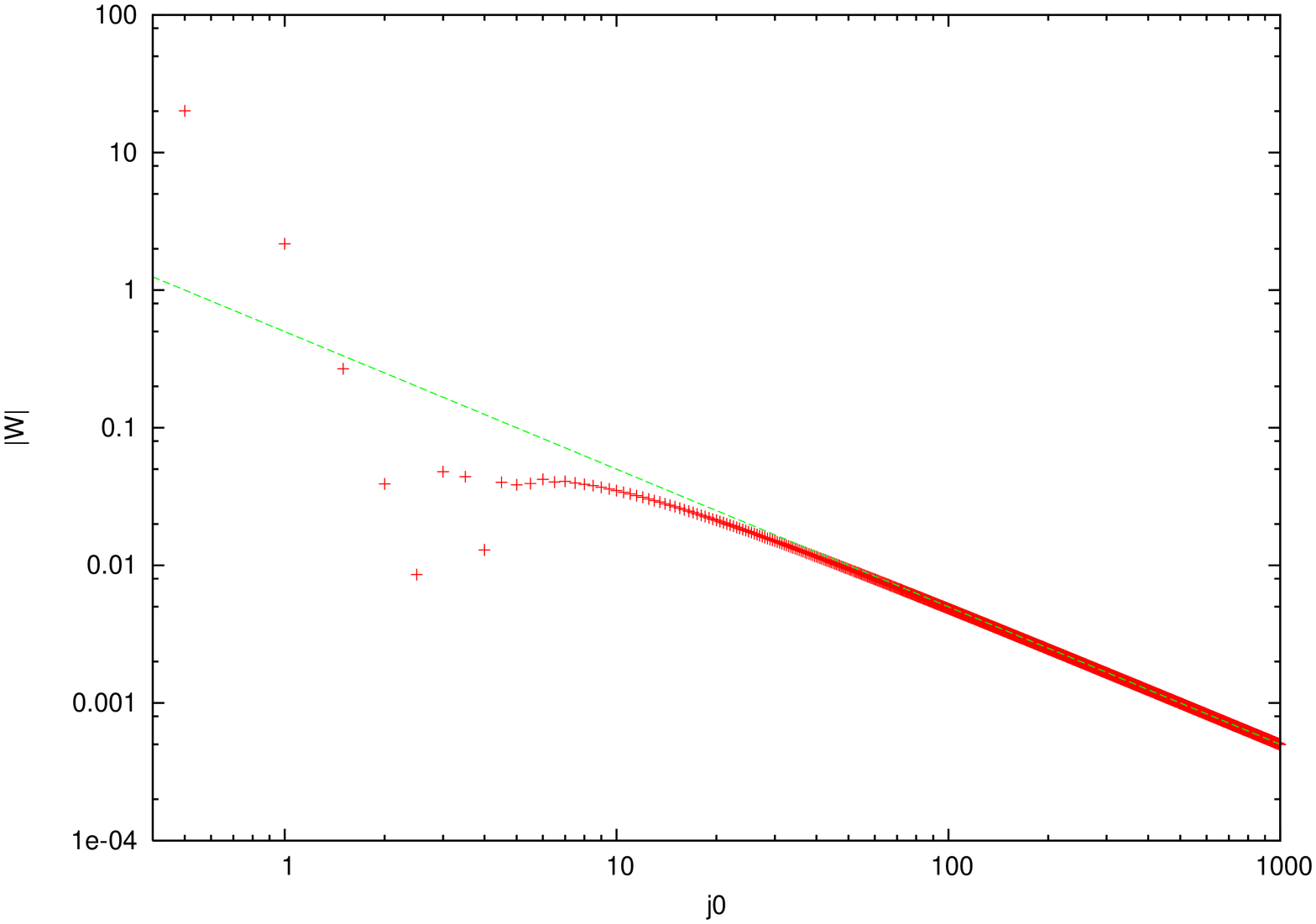}
\includegraphics[width=6cm,angle=-90]{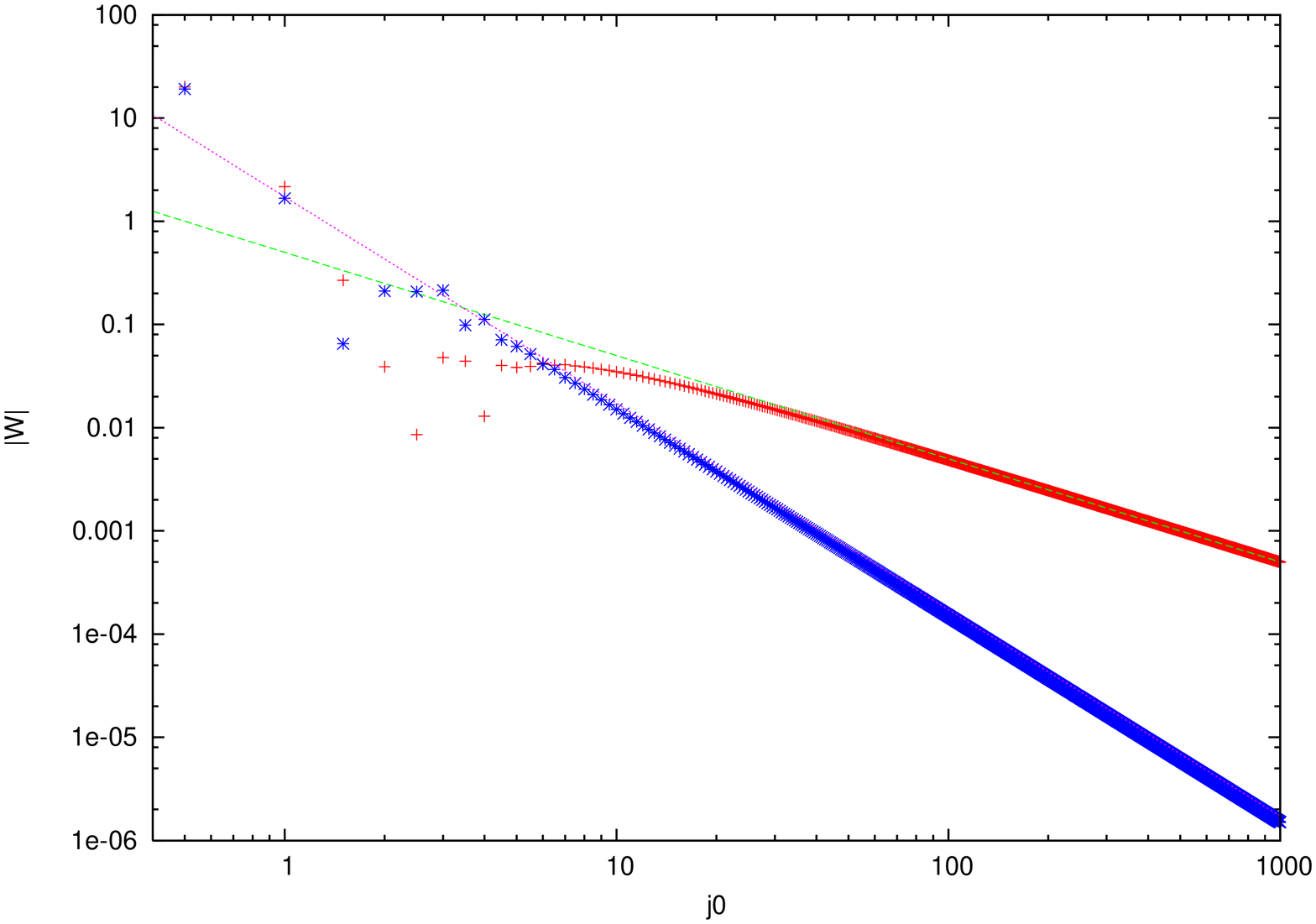}
\end{center}\caption{\small{The plots for the real boundary state
\Ref{real1}. Left plot: the leading order. In red, the numerical evaluations; the dashed green line is the analytic result in
 \Ref{Wsin}. Right plot: the numerical evaluations of the next to leading order, in blue,
 together with the analytic result of the next to leading in \Ref{Wsin}.}}\label{fitreal}
\end{figure}

The immediate choice to make the phase real is to take either the real or the imaginary part of \Ref{fase}, i.e.
$e^{i(j+\f12)\vartheta}\pm e^{-i(j+\f12)\vartheta}.$
In principle, it is not obvious that such a phase term will still guarantee the $1/j_0$
behaviour, thus we find it useful to address this issue within the toy model so far
described. To fix ideas, let us choose the real part, and consider a new boundary state
in
\Ref{Wexact} whose phase term is given by $\prod_i \cos (j_i+\f12)\vartheta$.
We keep the same quadratic term as in \Ref{psi0}, and we use again \Ref{asymp} to study the leading order.
This leads to a structure of the type
\equ
\left(e^{i(j_1+\f12)\vartheta}+e^{-i(j_1+\f12)\vartheta}\right)
\left(e^{i(j_2+\f12)\vartheta}+e^{-i(j_2+\f12)\vartheta}\right)
\Big(e^{iS_{\rm R}+i\f\pi4}+e^{-iS_{\rm R}-i\f\pi4}\Big).
\nequ
Recall that the linear order of the Regge action is $[4 (j_0+\f12)+ j_1+j_2+1 ]\vartheta$.
As discussed in Section \ref{SectionNLO}, only the exponential of the Regge action when
the linear term is cancelled will contribute to the 2-point function, the other term being
negligible due to the oscillations in the sum. It is thus clear that taking the cosine as the phase term makes in
principle both exponentials of the Regge action matter.
However, attention must be paid to the factor $4 (j_0+\f12)\vartheta$ in the linear order of the
Regge action.
In Section \ref{SectionNLO} we were able to discard the related phase
$e^{-i4(j_0+\f12)\vartheta}$, by reabsorbing it
in the normalisation. This can not be done now, because both orientations matter. Including
also the normalisation, we indeed have
\eqa\label{exp1}
W_{1122}(j_0)&=&\f1{j_0^4}\f{e^{-i4 (j_0+\f12)\vartheta-i\f\pi4}\left[b_1 j_0^5+ \ldots\right]
+ e^{i4 (j_0+\f12)\vartheta+i\f\pi4}\overline{\left[b_1 j_0^5+ \ldots\right]}}
{e^{-i4 (j_0+\f12)\vartheta-i\f\pi4}\left[B_1 j_0^2+ \ldots\right]
+ e^{i4 (j_0+\f12)\vartheta+i\f\pi4}\overline{\left[B_1 j_0^2+ \ldots\right]}}= \no \no &=&
\f1{j_0^4}\f{{\rm Re}\left[b_1 j_0^5+ \ldots\right]
+{\rm Im}{\left[b_1 j_0^5+ \ldots\right]}\tan{\left[4 (j_0+\f12)\vartheta+i\f\pi4\right]}}
{{\rm Re}\left[B_1 j_0^2+ \ldots\right]
+{\rm Im}{\left[B_1 j_0^2+ \ldots\right]}\tan{\left[4 (j_0+\f12)\vartheta+i\f\pi4\right]}}.
\neqa
While the first terms of numerator and denominator would produce the right $1/j_0$
behaviour, we see that we have an extra $j_0$-dependent phase which spoils the result.
Therefore, the choice $\prod_i \cos (j_i+\f12)\vartheta$ does not work. It is easy to see
that this problem can be fixed by taking
\equ\label{real1}
\Phi_\vartheta(j_i) = \cos(j_i+2j_0+\f32)\vartheta,
\nequ
so that the linear term of the Regge action is exactly cancelled.
In this way we get
\equ\label{pippo}
W_{1122}(j_0) = \f1{j_0^4}\f{{\rm Re}\left[b_1 e^{-i\f\pi4} j_0^5+ \ldots\right]}
{{\rm Re}\left[B_1 e^{-i\f\pi4} j_0^2+ \ldots\right]}
\nequ
Indeed, this latter choice cancels all $j_0$-dependent phases, and the $1/j_0$ behaviour
is restored. In particular, using the same quadratic term as in \Ref{psi0}, we obtain the
same $b_1$ and $B_1$ of \Ref{exp}, and thus the LO of \Ref{pippo} is 
\equ\label{Wsin}
W_{1122}(j_0)\,= \,\f{1}{j_0}\left(\f{{\rm Re} \, b_1+{\rm Im} \, b_1}{{\rm Re} \,
B_1+{\rm Im} \, B_1}\right)
\,=\,
\f{3\sin\f32\vartheta}{2j_0\sin\f\vartheta2}=\f1{2j_0}.
\nequ
This result is indeed supported by the numerics (see Fig.\ref{fitreal}).
Therefore, it is possible to keep the simple
diagonal and real quadratic term in the boundary state, and choose a phase term such that the
2-point function is a real quantity at all orders, with leading order proportional to $1/j_0$.

However, let us stress that this construction is not physically interesting in the
measured--time setting, where the 2-point function is a complex function, as mentioned
above. This is reflected in the fact that \Ref{real1} is not a sensible choice for the
phase term: the boundary is defined on the hyperplanes and it should not refer to the
bulk edges. On the other hand, the interest in the computation performed above is to have
an exploratory investigation towards the general boundary context. In fact, in the
general boundary setting, all edges $e$ of the tetrahedron are allowed to fluctuate, thus
it makes sense to take $\prod_e \cos (j_e+\f12)\vartheta$ as the phase term of the
boundary state. Since this quantity contains a phase that cancels exactly the linear
order of the Regge action, we expect its use to give a real $1/j_0$ behaviour. The
calculation performed here supports the idea that such a choice would give a correct real
2-point function. In the measured--time setting, the only interest in \Ref{real1} is to
show that using a real phase term would still work, with the benefit of a real 2-point
function.

We conclude with a remark about choosing the imaginary part of \Ref{fase}. Because of the
normalisation in \Ref{Wexact}, this is equivalent to take the $\SU(2)$ character itself, i.e.
\equ\label{phase}
\Phi_\vartheta(j)=\chi_{j}(\f\vartheta2) = \f{e^{i(j+\f12)\vartheta}-e^{-i(j+\f12)\vartheta}}{2i \sin\f\vartheta2}.
\nequ
It is interesting to note that \Ref{phase} can be interpreted as the propagator  of a point particle
coupled to the PR model \cite{etera}. Consequently, the use of \Ref{phase} as the phase term
of the boundary state opens the way to a new possible interpretation of
the gauge--fixing performed in this computation: \Ref{Wsin} can be thought of as the 2-point function
in the gauge fixed by the presence of a particle in the boundary.\footnote{With this interpretation,
it is natural to suggest that the dihedral angle $\vartheta$ is the deficit angle produced by the
particle on the boundary. But recall that the explicit value of the dihedral angles of the boundary
depends on the choice of triangulation, $\vartheta$ being the case of a single equilateral tetrahedron;
thus, this interpretation suggests a possible link between
boundary matter and finiteness of the triangulation \cite{inprep}.}

\subsection{Harmonic analysis}

We parametrise $\SU(2)$ elements as
$$
g(\phi, \hat n) = \cos\f\phi2 \, \mathbbm{1}+ i \, \sin\f\phi2 \, \hat n\cdot
\vec\sigma,
\quad \phi\in[0,4\pi[, \quad \hat{n}\in{\cal S}^2,
$$
where $\sigma_i$ are the Pauli matrices, satisfying $\sigma_i^2=\mathbbm{1}$. Moreover,
the group element $g(\phi, \hat n)$ is obviously identified to $g(-\phi, -\hat n)$. We
can thus restrict $\phi$ to live in $[0,2\pi]$. Finally, $g(\phi, \hat n)$ and
$g(\phi+2\pi,\hat n)\,=\, -g(\phi, \hat n)$ actually define the same rotation in the 3d
space (same $\SO(3)$ group element). In this representation, the characters and the Haar
measure are given respectively by:
$$
\chi_j(g)=\f{\sin(j+\f12)\phi}{\sin\f\phi2}, \qquad
dg=\f1{4\pi^2} \sin^2\f\phi2\, d^2\Omega(\hat n)\, d\phi.
$$
A class function $f$ is a function on the group which is invariant under conjugate
action, $f(g)= f(h g h^{-1})$. With the parametrisation chosen above, it is a function of
the angle $\phi$ only. For such a function, the Fourier transform can be simply written
as $f(g)= \sum_j \hat f_j \,
\chi_j(g)$, with $\hat f_j =\int dg\, f(g)\, \chi_j(g)$. If $j$ and $\phi$ in \Ref{psi0}
are conjugate variable like this, then just as \Ref{psi0} is a Gaussian peaked around
$j_0$ in the irrep labels basis, then its Fourier transform should describe a Gaussian on
the group peaked around the class element $\vartheta$. However, in spite of its simple
structure, \Ref{psi0} does not admit a simple Fourier transform. For, in the limit $j_0
\mapsto \infty$, we can approximate the sum with integrals, and find:
\eqa\label{Gauss2}
\widetilde\Psi_0(\phi)&=&\sum_j \Psi_0(j) \,{\chi_j(\phi)}=
\sum_j e^{-\f{\alpha}{2}(j-j_0)^2 + i(j+\f12)\vartheta} \,{\chi_j(\phi)} \no &\simeq&
\sqrt{\f\pi{2\alpha}}\f1{2i\sin\f\phi2}\,\left[
e^{i(j_0+\f12)(\vartheta+\phi)}e^{-\f1{2\alpha}(\phi+\vartheta)^2}
-e^{i(j_0+\f12)(\vartheta-\phi)}e^{-\f1{2\alpha}(\phi-\vartheta)^2}\right].
\neqa
However, notice that a function of the type $e^{-\f1{2\alpha}\phi^2}$
is not smooth (or even continuous) on
$\SU(2)$. This is due to the compactness of $\SU(2)$ and the gluing condition at the
``boundary'' $\phi\,=\,0\,[2\pi]$.

\begin{figure}[t]
\begin{center}\includegraphics[width=6cm,angle=-90]{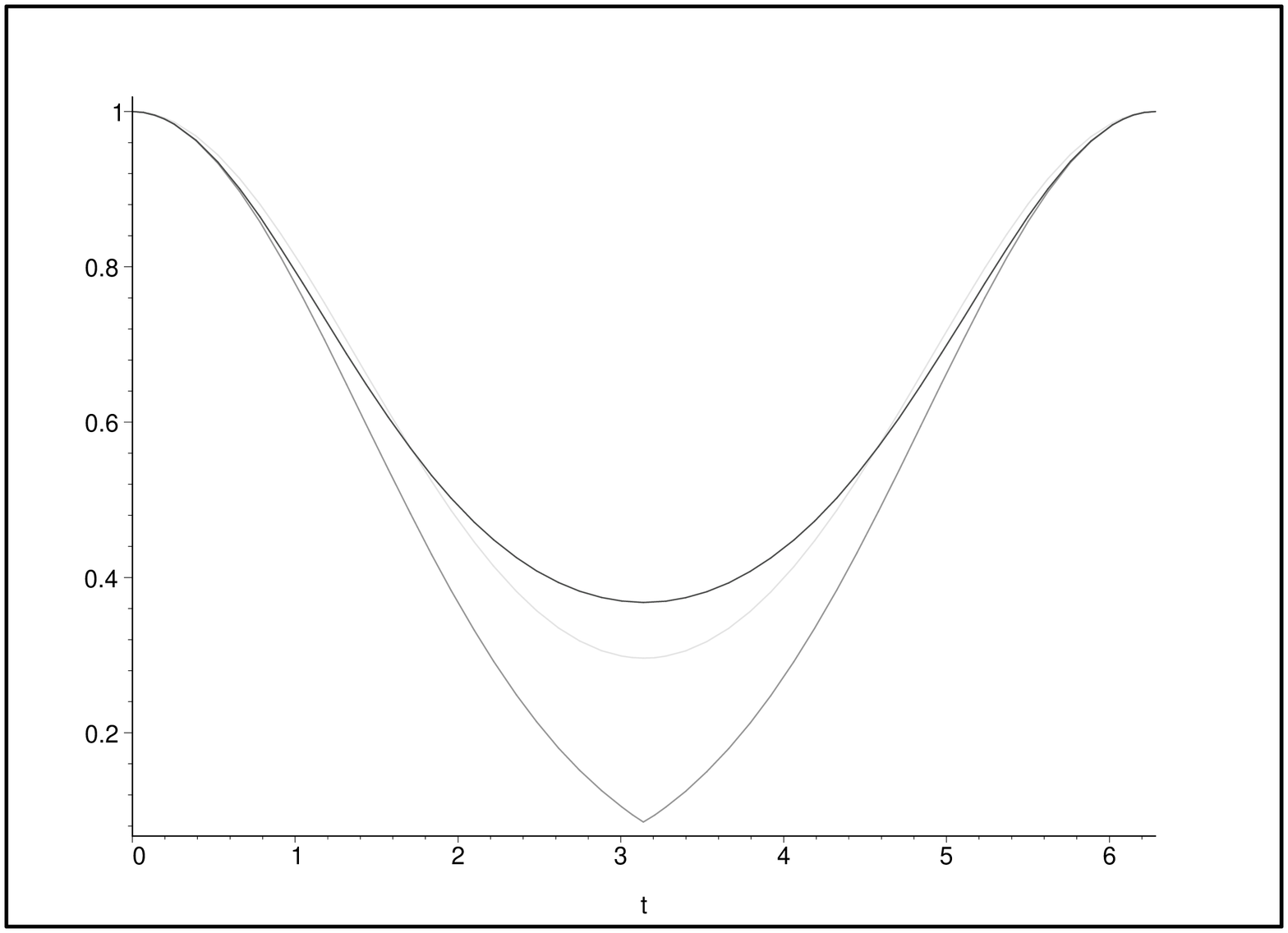}
\includegraphics[width=6cm,angle=-90]{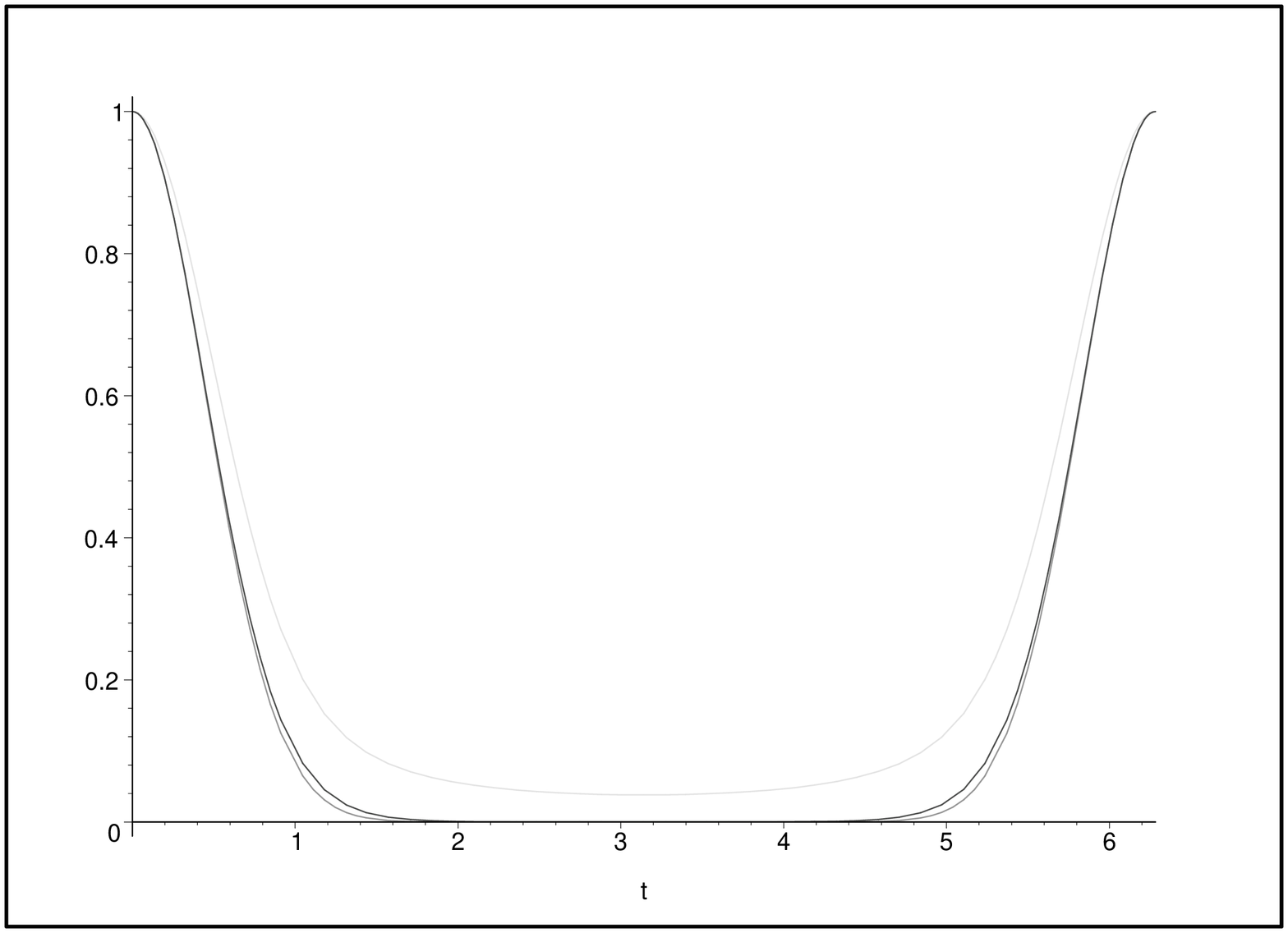}
\end{center}
\caption{\small{
We plot the three considered Gaussian distributions on $\SU(2)$ peaked around
$\pm\mathbbm{1}$: $f_1(\phi)=e^{-\f{2}{\alpha}\sin^2\f \phi2}$, $f_2(\phi)=\sum_j
\chi_j(\phi)e^{-\f\alpha2 j^2}$ and $f_3(\phi)=e^{-\f{1}{2\alpha}\phi^2}$
(for $\phi\in[0,\pi]$ and its mirror image $e^{-\f{1}{2\alpha}(2\pi-\phi)^2}$ for
$\phi\in[\pi,2\pi]$). The two plots respectively correspond to the values $\alpha=2$ and
$\alpha=0.2$. On the first plot, it is clear that the three Gaussians are distinct and
that $f_3$ is not smooth. On the second, we see that $f_2$ converges toward the
distribution $f_3$ as $\alpha$ goes to 0 (high $j_0$), thus justifying \Ref{Gauss2} while the distribution $f_1$ dual
to the Bessel functions remains slightly different. This deviation of $f_1$ from the
standard $f_2$ will be responsible for the 
different NLO of the 2-point function when using
it as boundary state (see below).}}
\label{GaussianPlots}
\end{figure}

A smooth, more natural Gaussian--like function on $\SU(2)$ would be 
\equ\label{Gauss4}
\widetilde\psi_0(g) = \f1{N'} e^{\f1{2\alpha}{\rm Tr} (g\cdot\vec\sigma)^2}
= \f1{N'}e^{-\f2\alpha \sin^2\f\phi2},
\nequ
where the trace Tr is taken in the fundamental $j=1/2$ irrep. This function can be easily
decomposed into its Fourier modes; notice in fact that ${\rm Tr}
(g\cdot\vec\sigma)^2=\chi_{\f12}^2(g)-4$, thus \Ref{Gauss4} can be expanded in a simple
sum of characters, thanks to the decomposition property of tensor products into irreps,
$\chi_{\f12}^n=\sum_j c_j \chi_j$. This feature makes it a more natural object to use
within the group structure of $\SU(2)$. For a comparison with \Ref{Gauss2}, see Fig.\ref{GaussianPlots}.

To find the normalisation $N'$, we evaluate the integral
\equ\label{norm}
|N'|^2= \int_{\SU(2)} dg \,e^{\f1{\alpha}{\rm Tr} (g\cdot\vec\sigma)^2}=
\f1\pi\int_0^{2\pi} d\phi \, \sin^2\f\phi2 \, e^{-\f4{\alpha} \sin^2\f\phi2}=
e^{-\f2{\alpha}} \left[ I_{0}(\f2{\alpha}) - I_{1}(\f2{\alpha})\right],
\nequ
where $I_n(\f1\alpha)$, $n\in \mathbbm N$,
are the modified Bessel functions of the first kind (see the Appendix).
In a similar way, we can compute the Fourier transform of \Ref{Gauss4}. For
$j\in \mathbbm N$, we have
\equ\label{bessel}
\psi_0(j) = \int_{\SU(2)} dg \,\widetilde\psi_0(g) \, \chi_j(g) =
\f1{N'} \f1\pi \int_0^{2\pi} d\phi \, \sin\f\phi2 \, \sin(2j+1)\f\phi2 \,e^{-\f2{\alpha} \sin^2\f\phi2}
=\f{ I_{j}(\f1{\alpha}) - I_{j+1}(\f1{\alpha})}
{\sqrt{I_{0}(\f2{\alpha}) - I_{1}(\f2{\alpha})}},
\nequ
where we used the explicit value \Ref{norm}. On the other hand, $\psi_0(j)$ vanishes if $j$ is strictly an
half--integer.

At first sight, it is not obvious that \Ref{bessel}  fulfills the semiclassical
properties required earlier. However, as we show in the Appendix, the $\alpha\mapsto 0$
limit of \Ref{bessel} is $\sqrt{2}\sqrt[4]{\f\alpha\pi}e^{-\f\alpha2j^2}$, namely a
Gaussian peaked around $j=0$. We are on the right track. To obtain a function peaked
around $j_0$, we need to shift the peak. This can be done in a simple way, now that we
understand the Gaussian in the irrep label basis as the Fourier transform of
\Ref{Gauss4}; to shift the peak of the Gaussian to a value $j_0$, it is enough to
multiply \Ref{Gauss4} by the appropriate Fourier mode, namely $\chi_{j_0}(g)$.
Accordingly, we consider the following function,
\equ\label{Gauss5}
\widetilde\psi_0(g)=
\f1{N} e^{\f1{2\alpha}{\rm Tr} (g\cdot\vec\sigma)^2} \chi_{j_0}(g),
\qquad N=e^{-\f1{\alpha}} \sqrt{ I_{0}(\f2{\alpha}) - I_{2j_0+1}(\f2{\alpha})}.
\nequ
Proceeding as above and using $\chi_j(g) \, \chi_{j_0}(g) = \sum_k \chi_k(g)$,
it is straightforward to check that the Fourier transform
of \Ref{Gauss5} is
\equ\label{bessel2}
\psi_0(j) = \f1{N}  \sum_{k=|j-j_0|}^{j+j_0}
\int_{\SU(2)} dg \, e^{\f1{2\alpha}{\rm Tr} (g\cdot\vec\sigma)^2}
\chi_k(g) =
\f{ I_{|j-j_0|}(\f1{\alpha}) - I_{j+j_0+1}(\f1{\alpha})}{\sqrt{I_{0}(\f2{\alpha}) - I_{2j_0+1}(\f2{\alpha})}}.
\nequ
Again, this result holds for $j$ integer, whereas $\psi_0(j)$ vanishes for $j$ strictly an half--integer.
For small $\alpha$, this is approximated by (see the Appendix)
\equ\label{BesselGauss}
\psi_0(j) \simeq \f{ I_{|j-j_0|}(\f1{\alpha})}
{\sqrt{I_{0}(\f2{\alpha})}} \simeq \sqrt{2} \sqrt[4]{\f\alpha\pi}\, e^{-\f\alpha2 (j-j_0)^2},
\nequ
and it describes a Gaussian peaked around $j_0$, with squared width $\f1{\alpha}$, as we wanted.

Let us now discuss the dihedral angle. \Ref{Gauss5} is a Gaussian peaked around $\phi=0$.
We want to shift this peak as well, around $\phi=\vartheta$. This can be done adding a phase in $j$.
The simplest way to do so, without spoiling the property of having a simple Fourier transform,
is to add to \Ref{bessel2} the real phase term $\cos(j+\f12)\vartheta$,
which we discussed in the previous section. Then, we consider
\equ\label{psicool}
\Psi_0(j) =
\f{ I_{|j-j_0|}(\f1{\alpha}) - I_{j+j_0+1}(\f1{\alpha})}{\sqrt{I_{0}(\f2{\alpha}) - I_{2j_0+1}(\f2{\alpha})}}
\, \cos (j+\f12)\vartheta.
\nequ
The reason why the cosine is the easier choice to add a phase in $j$ is that using
\equ\label{trigo}
\cos\left[\left(j+\f12\right)\vartheta\right] \, \chi_j(\phi)\,=\,
\f{\sin\f{\phi-\vartheta}{2}}{2\sin\f\phi2}\,\chi_j(\phi-\vartheta)\,+\,
\f{\sin\f{\phi+\vartheta}{2}}{2\sin\f\phi2}\,\chi_j(\phi+\vartheta)
\nequ
we can simply compute the Fourier transform of \Ref{psicool} shifting the
 character's argument as in \Ref{trigo}, and we get
\equ\label{psicool2}
\widetilde\Psi_0(g) \,=\,
\f{1}{2N\sin\f\phi2}\left[\,\sin(\f{\phi-\vartheta}{2})\,\chi_{j_0}(\phi-\vartheta)\,
e^{-\f2{\alpha}\sin^2\f{\phi-\vartheta}{2}} \,+\,
\sin(\f{\phi+\vartheta}{2})\,\chi_{j_0}(\phi+\vartheta)\,
e^{-\f2{\alpha}\sin^2\f{\phi+\vartheta}{2}}\,\right].
\nequ

Let us comment on this expression. In the general boundary context, \Ref{psicool} can be
taken as a new ansatz for the boundary state. It has a good semiclassical behaviour, in
the sense described above: in the $j_0\mapsto\infty$ limit, it becomes a Gaussian peaked
around $j=j_0$ (see \Ref{psicool}) and $\phi=\vartheta, \pi-\vartheta$ (see
\Ref{psicool2}). Furthermore, the interpretation of $j_0$ and $\vartheta$ as conjugate
variables has now a clear mathematical meaning: they are conjugate with respect to the
$\SU(2)$ Fourier transform.

Notice that we are led to include also the configuration $\phi=\pi-\vartheta$: the
$\SU(2)$ angle $\phi$ represents both the external dihedral angle $\vartheta$ and the internal angle $\pi-\vartheta$
between the triangles. In other words, the boundary spin network has an inside--outside symmetry.

\begin{figure}[ht]
\begin{center}\includegraphics[width=6cm,angle=-90]{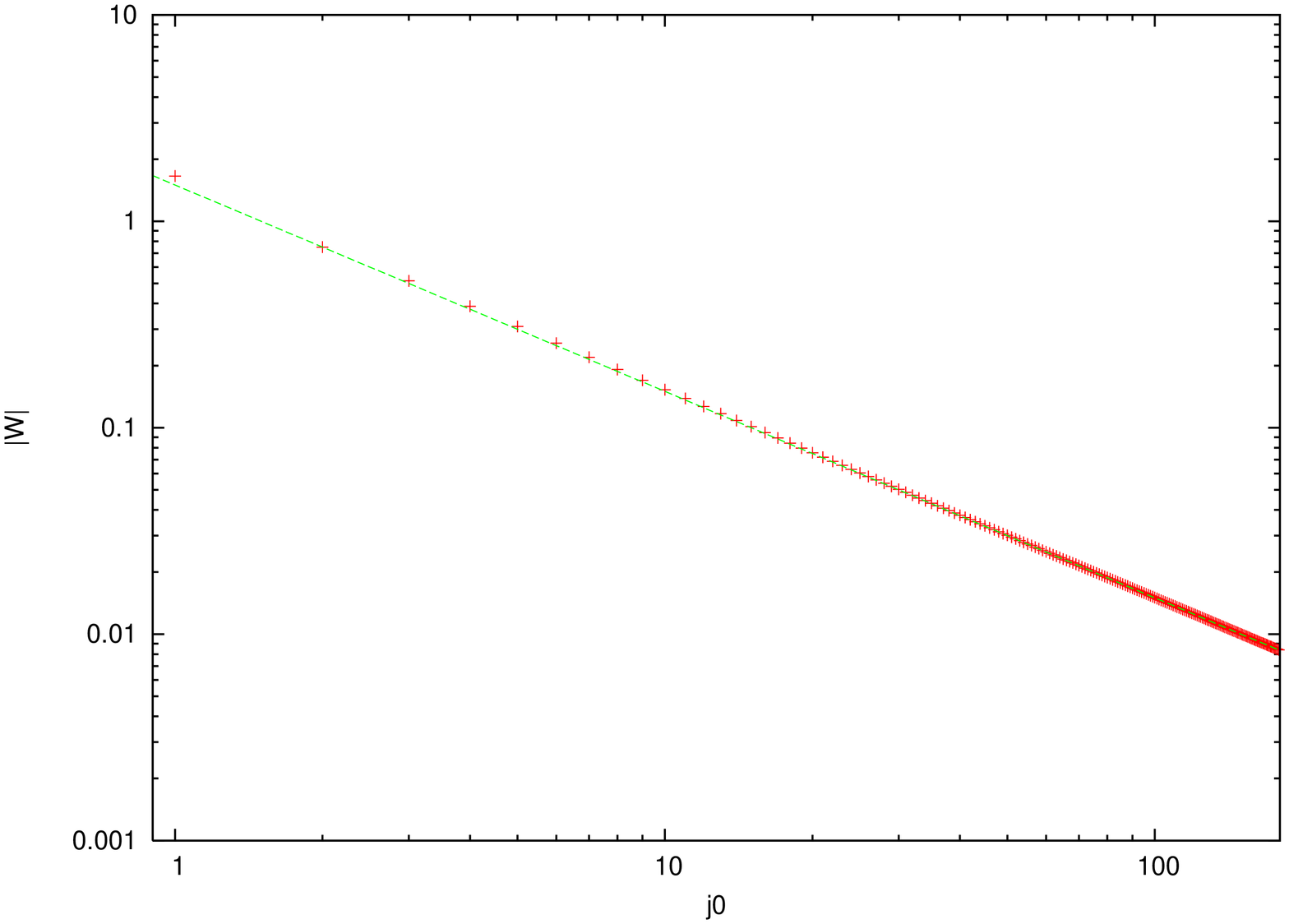}
\includegraphics[width=6cm,angle=-90]{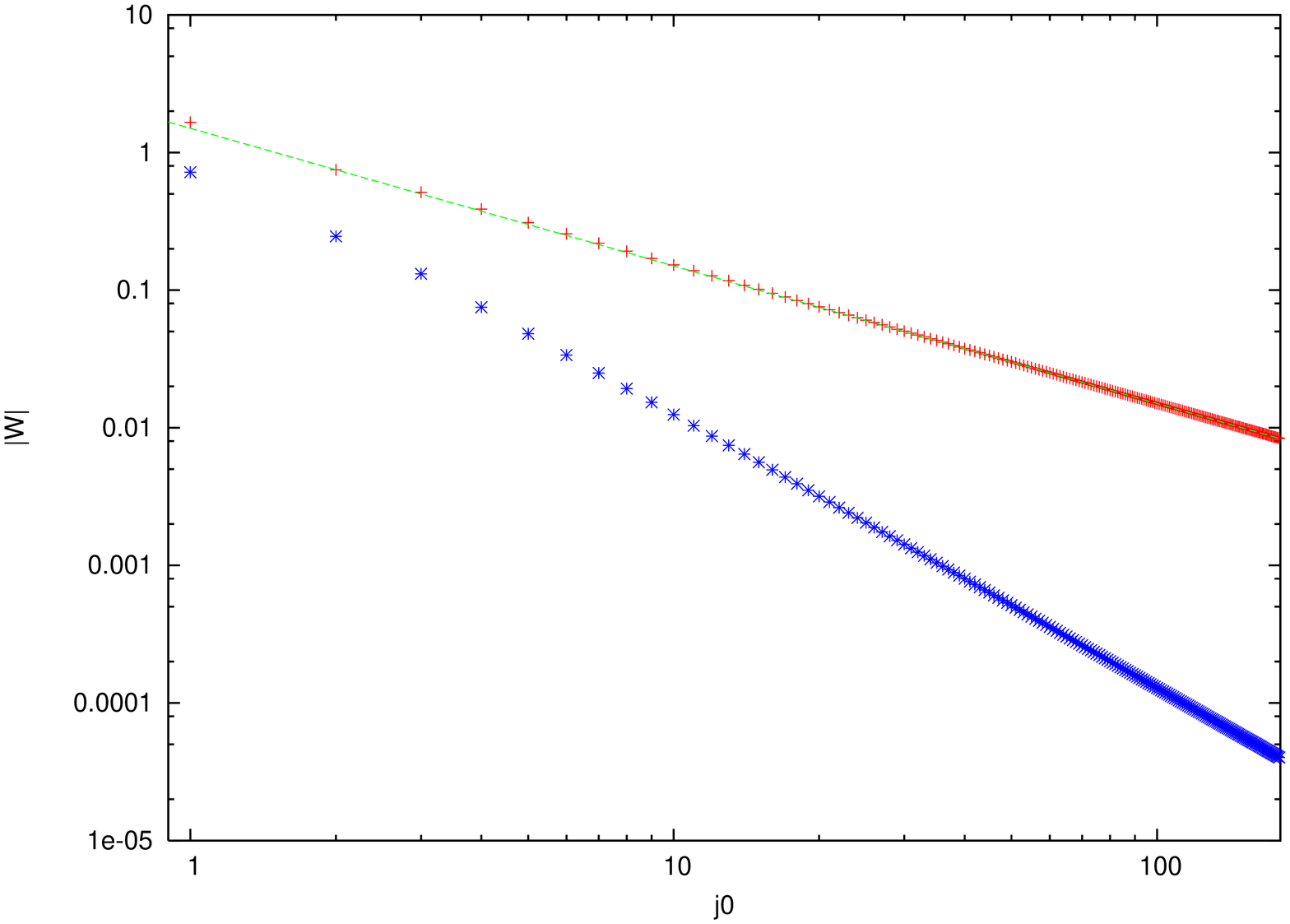}
\end{center}\caption{\small{Leading and next to leading orders using \Ref{bessel2}.
Left plot: the leading order. In red, the numerical evaluation; the dashed green line is the analytic result in
 \Ref{zero}. Right plot: the numerical evaluations of the next to leading order, in blue,
 together with the leading order plot.}}\label{besselfit}
\end{figure}

In the general boundary context, we expect the use of \Ref{psicool2} to lead to a real propagator
$W$ with the correct $1/j_0$ leading order.
This is suggested by the asymptotics \Ref{BesselGauss}.
To support this expectation, we consider this new ansatz
in the toy model presented in this paper. In our context however, we cannot use the cosine phase term,
as we discussed above. Then, we consider \Ref{bessel} for the quadratic term and the
the usual phase \Ref{fase}.
Choosing the same $\alpha=\f4{3j_0}$ as before,
we obtain from the numerics exactly the same leading order \Ref{LO} (see also Fig.\ref{besselfit}),
\equ\label{besselW}
W_{1122}(j_0)\simeq \f1{j_0}(-0.5+i1.4) + \f1{j_0^2}(-1.3-i0.2)
\nequ
Notice that the next to leading order is now different.\footnote{
The reason for fewer digits of precision lies in the numerical
difficulty of calculating the Bessel functions.}
This is due to the fact that \Ref{bessel2} contributes non trivial
corrections. Indeed, the leading order of its perturbative expansion, \Ref{BesselGauss},
reduces to the quadratic part of \Ref{psi0}, but higher order terms enter the calculation
of the propagator corrections.
From this point of view, the use of \Ref{bessel2} enriches the toy model, and it makes it
possible also to consider the most complete structure of the graviton perturbative
expansion, where both the kernel and the boundary state are given as a power series in $\lp$.

As for the contribution of these new corrections, notice that the NLO in \Ref{besselW} is
slightly smaller than the NLO \Ref{NLOn}. This might look like a side result. However, in
4d the numerical complexity grows very fast, a fact that has slowed down numerical
computations thus far. Therefore, having a new boundary state which reaches the
same asymptotic behaviour earlier could turn out to be very useful,
were the same behaviour to hold in 4d. Furthermore, it turns out that
using such a boundary state allows an exact calculation of the LO of the 2-point function
in the 4d case; namely, one does not need to use the (troubled) asymptotic formula of the
$\{10j\}$ symbol, but one can uses its exact expression in terms of integrals of
character. A work on this is in preparation \cite{noi2}.

\section{Comparing different spinfoam models: stability of the leading order}\label{sectionk}

The definition of the $n$-point functions in the spinfoam formalism
is not only fundamental to extract physical predictions, it is also
important for distinguishing between the different existing models.
In particular, obtaining the correct LO for the 2-point function
is a requirement for any sensible model.
In 4d, there are a variety of different models, and using the
equivalent of \Ref{Wexact}
to study their large scale limit could possibly be a good way to narrow them down.
However, some variety is also present for the 3d models, and it is useful to
consider the 3d toy model to begin to investigate how different models
can produce different 2-point functions.

We consider a simple generalisation of the PR model, where
we modify the face weights, by taking
\equ\label{Zk}
{\cal A}^k_\Delta(j_e) = \prod_e (2 j_e+1)^k \prod_\tau \left\{6j\right\}_\tau
\nequ
as the amplitude.
The original model is reproduced for $k=1$.

From the point of view of the Regge path integral \Ref{Wbis},
this new definition of the kernel amounts simply to a change in the measure term
\Ref{measure}, which becomes
\equ
\mu_k(j_e):=\f{\prod_i \, (2j_i+1)^k}{\sqrt{V(j_1,j_2,j_0)}}.
\nequ
If we take into account this change, we can straighforwardly
recompute the LO and the NLO, and we find that the $k$-dependence enters in a very simple way:
\equ
W^k_{1122}(j_0) = -\f1{j_0} (\f12-i\sqrt2)
-\f1{j_0^2} (\f12-i\sqrt2) \f12 [k^2-\f53k+\f{2161}{1296}+i(\f{7}{3\sqrt2}k+\f{17}{648\sqrt{2}})].
\nequ
Let us comment on this result. Firstly, the LO is insensitive to this
kind of modification. This means that any spinfoam model of the type \Ref{Zk} leads
to the same asymptotic behaviour of the 2-point function. This result is expected,
as we know from the previous analysis that at leading order only the trivial background
value for the measure enters, so that any $k$-dependence is washed away by the normalisation in \Ref{Wexact}.

\begin{figure}[ht]
\begin{center}\includegraphics[width=5cm,angle=-90]{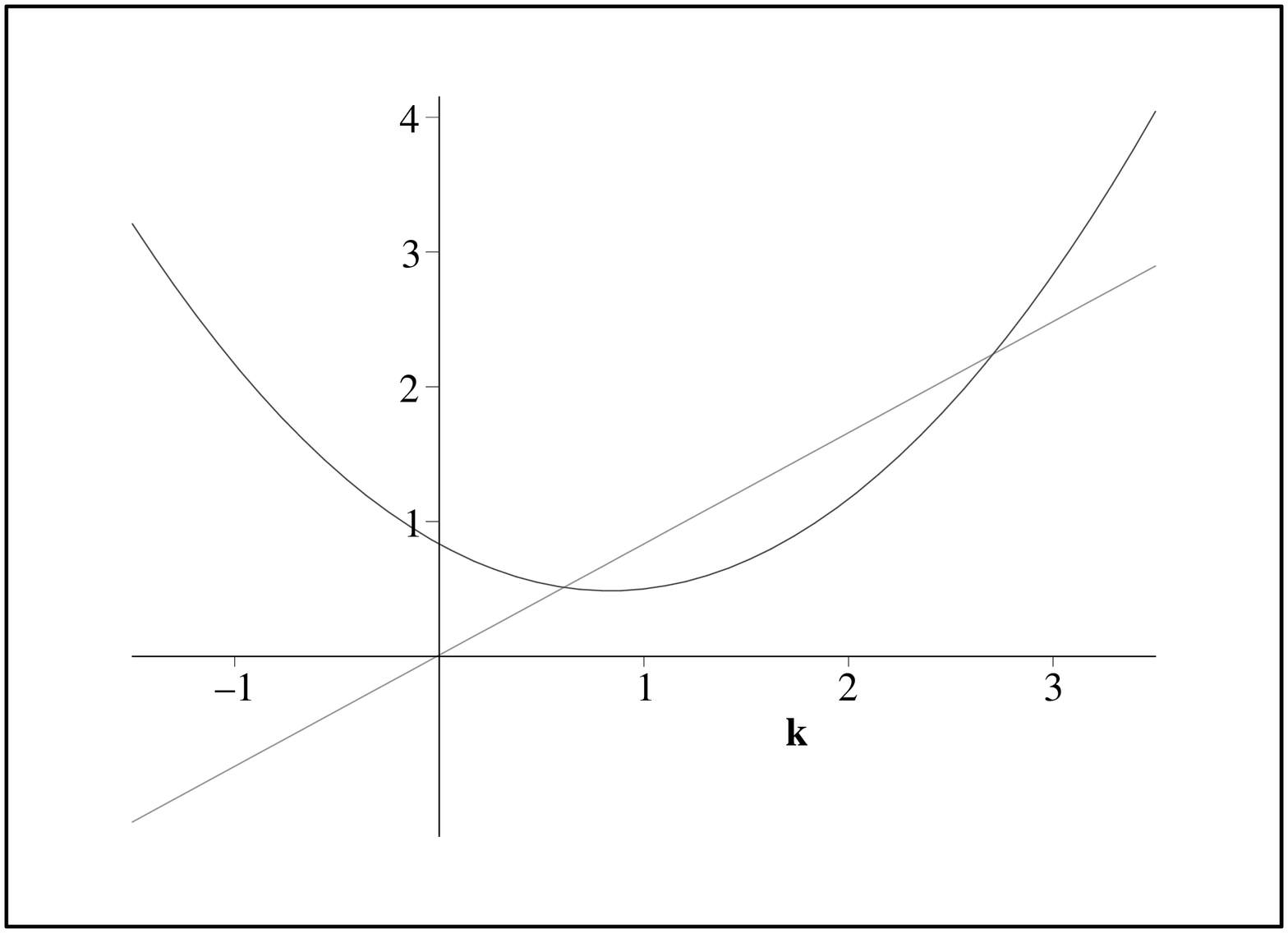}
\includegraphics[width=5cm,angle=-90]{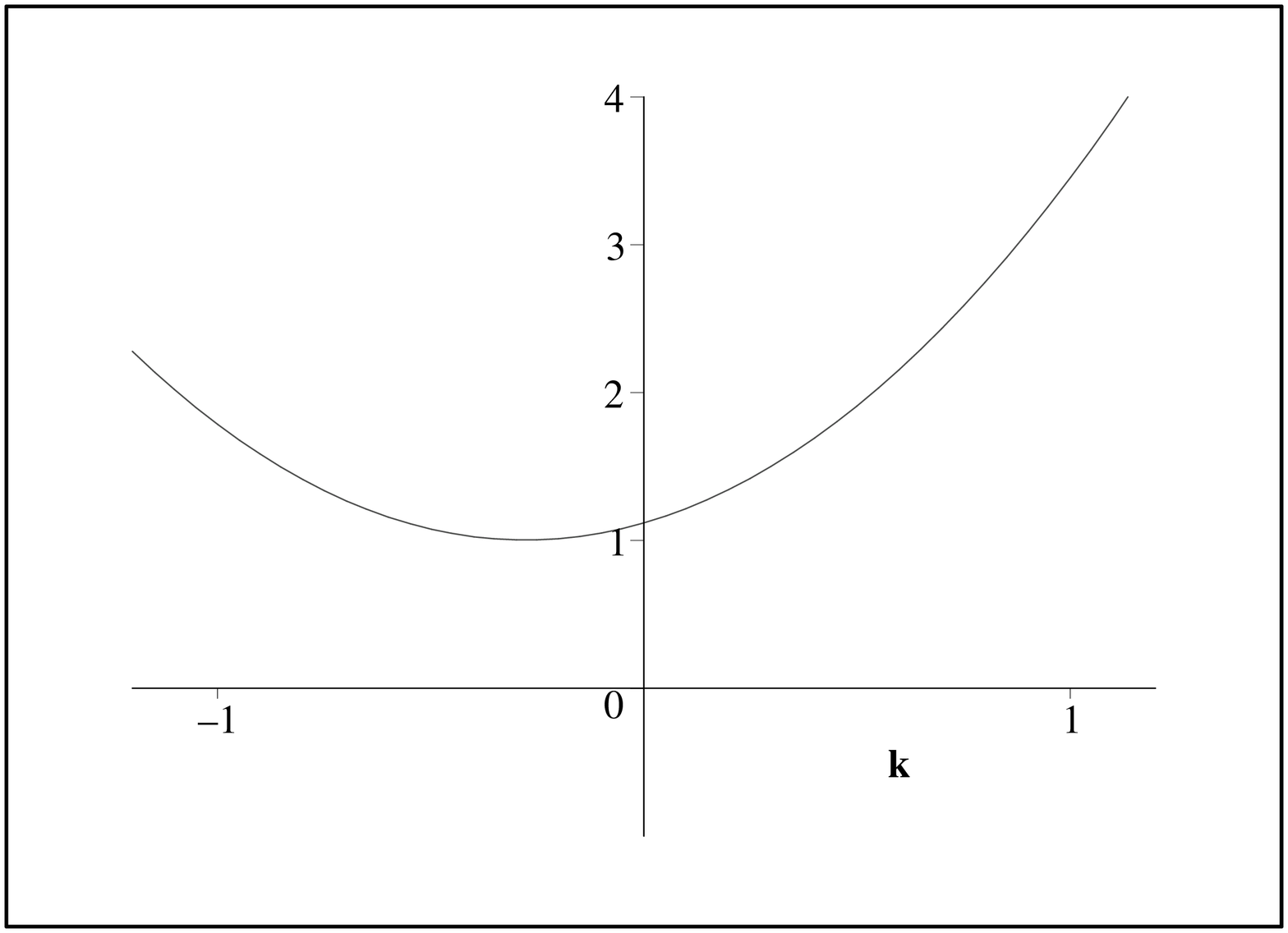}
\end{center}\caption{\small{The relative correction $W^{\scr NLO}/W^{\scr LO}$,
as a function of the measure weight $k$. On the left, the case with boundary state
\Ref{psi0}; the
quadratic function is the real part, while the linear function is the
imaginary part. We see that $k=1$ is the integer value closest to the minimum
of the real part. On the right, the real case with \Ref{real1} as the phase of
the boundary state. We see that $k=0$ is the integer value closest to the minimum.}}\label{k}
\end{figure}

On the other hand, the NLO depends on $k$, thus different models lead to different corrections.
While the (conventional) case $k=1$ remarkably minimises the real part of the relative
correction $W^{\scr NLO}/W^{\scr LO}$ (see left plot of Fig.\ref{k}), the minimum of
the absolute correction is reached for $k=0$,
the case of trivial face weights.\footnote{Interestingly,
it has been advocated in the 4d case that a model with trivial face weights could be better behaved \cite{Baez}.}
This feature is clarified if we use \Ref{phase} as the phase of the boundary state,
which leads to a real 2-point funtion. In this latter case,
the $k$-dependence of the correction is
\equ
W^k_{1122}(j_0) = -\f1{2j_0}
-\f1{j_0^2} (\f34k^2+\f5{12}k+\f{8701}{15552}),
\nequ
and we see that the case $k=0$ minimises the correction (see the right plot of Fig.\ref{k}).

\section{Extending the toy model: the isosceles case}\label{sectionrett}

In this last section, we want to address a different issue.
In the toy model considered so far, we have restricted our attention
to the case when the background is chosen to be an equilateral tetrahedron.
However, we expect the results here described to extend to a generic background
configuration. To give an idea of the way things would work out,
and as a first generalisation, we consider in this Section the case
when the two opposite edges of the background tetrahedron have spin $j_0$,
and the four bulk edges have spin $j_t$. We call the corresponding dihedral angles
$\vartheta_0$, $\vartheta_t$ (all relevant formulas are reported in the Appendix).
In this case,
we can still consider an embedding in 3d Euclidean spacetime as in
Fig.\ref{planes}, and $T=t_2-t_1=\f1{\sqrt 2}{\sqrt{2j_t^2-j_0^2}}$.

The physical setting of the problem is exactly the same. The only
difference is the geometric analysis, which is now slightly more involved.
In particular, notice that not all configurations $(j_0, j_t)$ are admissable;
since the tetrahedron is embedded in Euclidean spacetime, triangle inequalities must hold,
and thus we restrict in the following to $\sqrt2j_t\geq j_0$.

Using the simple ansatz of the type \Ref{psi0} for the boundary state,
the expression \Ref{Wexact} for the propagator now reads
\equ\label{Wrett}
W_{1122}(T) = \f1{j_0^4}\,\f1{\cal N}\, \sum_{j_1, j_2=0}^{2j_t} \left[\prod_i \, (2j_i+1)
\;[C^2(j_i) - C^2(j_0)] \,\Psi_0[j_i]\right]
\left\{ \begin{array}{ccc} j_1 & j_t & j_t \\ j_2 & j_t & j_t \end{array} \right\},
\nequ
with
\equ\label{psi0rett}
\Psi_0[j_i]= \exp\left\{ -\f\alpha2(j_i-j_0)^2+i\vartheta_0(j_i+\f12) \right\}
\nequ
and $\cos\vartheta_0 = -\f{4j_t^2-3j_0^2}{4j_t^2-j_0^2}$ (see the Appendix).
To obtain the leading order of \Ref{Wrett}, we study the limit $(j_t,j_0)\mapsto\infty$
with $r=j_t/j_0$ fixed. This last requirement is necessary in order for \Ref{asymp}
to hold; furthermore, it would not be very interesting to study
the inhomogeneous limit $j_t\mapsto\infty$ with $j_0$ fixed, because this
means going towards a degenerate configuration where the tetrahedron is squeezed
into a 2d surface. In the limit $(j_t,j_0)\mapsto\infty$,
we proceed as in Section \ref{SectionNLO}, using \Ref{asymp} for the \6 symbol,
and expanding the Regge action.
Choosing the value
\equ\label{alpha}
\alpha=\f{4 r}{j_0 (4 r^2-1)}
\nequ
for the free parameter entering \Ref{psi0rett}, we obtain
\equ\label{Wrett2}
W_{1122}(j_t, j_0) = \f1{j_0^4}\,\f1{\cal N} \sum_{j_1, j_2}
F(j_0, \d j_1, \d j_2) \;e^{ -\f{1}{2} \sum A_{ik} \d j_i \d j_{k}}
\nequ
with
\equ
A_{ik} = \f{j_0 (4 r^2-1)}{4 r} \left( \begin{array}{cc} 1+i \cot\vartheta_t & -\f i{\sin\vartheta_t} \\ \\
-\f i{\sin\vartheta_t} & 1+i \cot\vartheta_t \end{array}
\right).
\nequ
Here we used the fact that $\cos\vartheta_t=-\f1{4r^2-1}$ (see the Appendix).
Once again, this matrix can be interpreted as the kernel for a harmonic oscillator
(see for instance \cite{carlo}, or the appendix of \cite{Io}), if we
identify the frequency of the oscillator with
$\om = \f{\vartheta_t}{T}.$ Indeed, it is this request that motivated our choice \Ref{alpha}.

At leading order, $F(j_0, \d j_1, \d j_2)$ can be read from the first term of \Ref{Fexpanded};
the analogous term in the normalisation is the first term of \Ref{FNexpanded}.
Approximating the sums as Gaussian integrals as in Section
\ref{SectionNLO}, we get
\equ\label{LOrett}
W_{1122}(j_t, j_0) = 
\f{4r^2-1}{2j_t} e^{i\vartheta_t}.
\nequ
For $r=1$, we recover the equilateral result \Ref{LO}. To consider non equilateral configuration,
we plot \Ref{LOrett} for different values of $r$.
\begin{figure}[ht]
\begin{center}
\includegraphics[width=6cm,angle=-90]{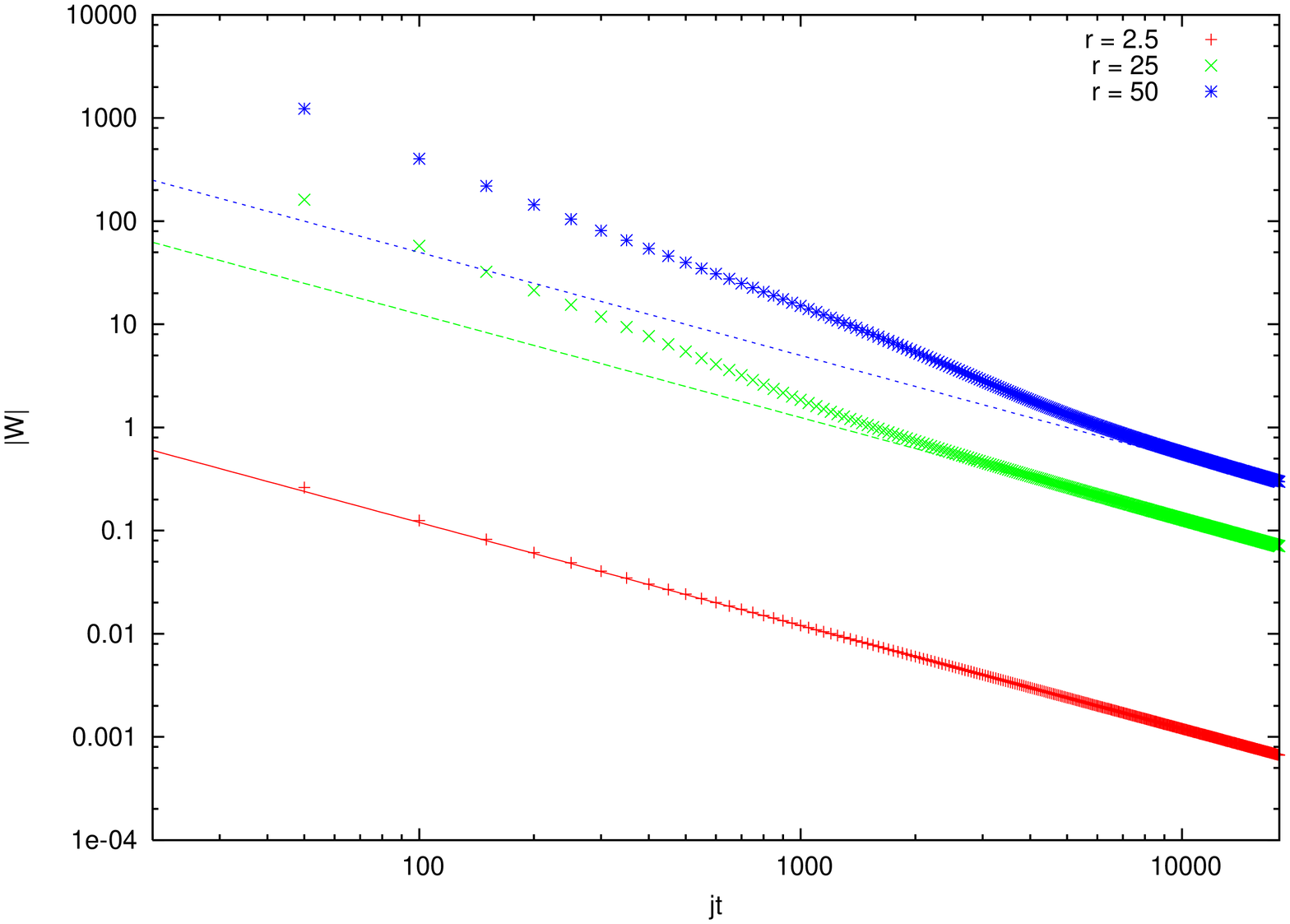}
\includegraphics[width=6cm,angle=-90]{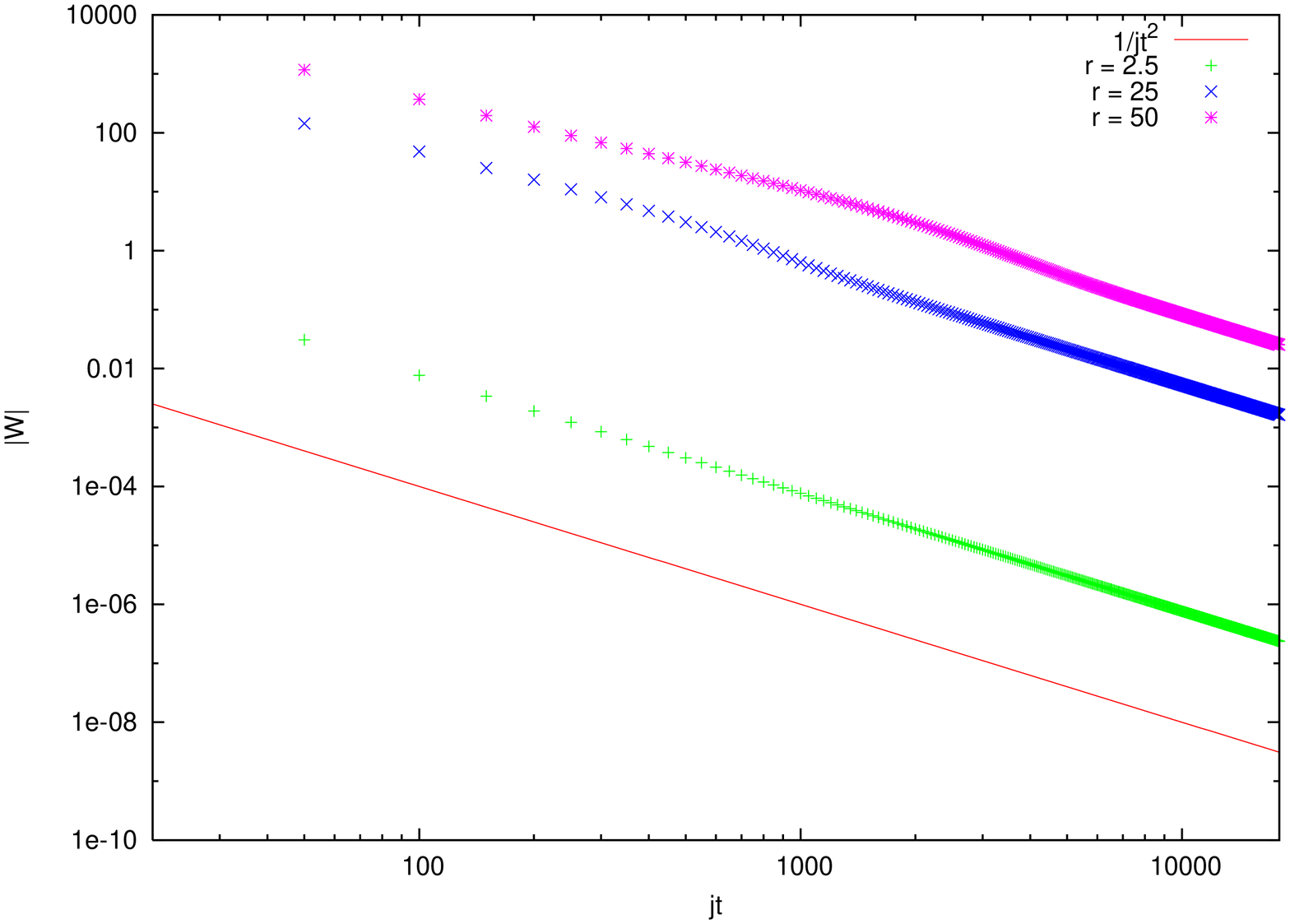}
\end{center}\caption{\small{Exact numerical evaluation of $|W_{1122}|$
  as  a function of $j_t$, for $r = 2.5, 25,$ and $50$. On the left, the 
  leading order, computed in \Ref{LOrett}. On the right, the next to leading order, not computed
  analytically here.}}\label{boh}
\end{figure}
In particular, if $r\gg1$, we have $\cos\vartheta_t\simeq\f\pi2$ and $\om\simeq\f\pi{2r j_0}$,
and \Ref{LOrett} reads
\equ\label{LOrettlimit}
W_{1122}(T) \simeq \f{2\pi}{j_0^2} \f{e^{i\om T}}{2\om}.
\nequ

A comparison of the exact numeric calculation with the asymptotics is
shown in Figure~\ref{boh}, where we see good agreement with the
prediction of~\Ref{LOrett}.  Note, however, that the asymptotic
value is reached more and more slowly as $r$ increases.

\section{Conclusions and perspectives}

We have considered a single tetrahedron embedded in 3d Euclidean spacetime as a toy model to
test how the spinfoam formalism can be used to construct the perturbative quantum theory
of gravitons. We focused on a single component of the propagator, defined in
\Ref{Wexact}, and computed the NLO and NNLO corrections to the free $1/\ell$ behaviour.
These are given respectively in \Ref{NLOn} and \Ref{NNLOn} and, as expected, have the
behaviour $1/\ell^2$ and $1/\ell^3$.

We found that these corrections have an interesting structure. On the one hand, the NLO
can be entirely computed from an expression, given in \Ref{Wbis}, which is a path
integral for the Regge action. This can be seen as the discrete analogue of the
conventional continuum definition of the 2-point function. On the other hand, the NNLO
gets contributions from the true, non Regge--like spinfoam dynamics. This in turn cannot
be written down by simply discretizing the GR action \emph{\`a la} Regge. Here is where
the use of the spinfoam formalism modifies substantially the conventional expansion. It
is particularly intriguing that this new source of corrections enters at NNLO, precisely
where the continuum theory becomes non renormalisable, and that it
reduces the magnitude of the NNLO correction. 

We introduced a new ansatz for the boundary state, given in \Ref{psicool}, which reduces to 
the original Gaussian in the large $j$ limit, and thus satisfies the semiclassical
requirements, but has the advantage of making $j$ and the dihedral angle conjugate variables
with respect to the harmonic analysis over $\SU(2)$. This property helps treating the sums
appearing in the evaluation of expectation values within the PR model. 
In particular in 4d, the Riemannian Barrett--Crane model uses the gauge group
$\Spin(4)\sim \SU(2)\times\SU(2)$, to which we can apply all the tools developed here for
the gauge group $\SU(2)$. Remarkably, the use of \Ref{psicool} in the general boundary context
allows us to write the 4d propagator as an integral over the group \cite{noi2}.

It would be interesting to extend these results to the Lorentzian case, using 
the expression \Ref{Wexact} with group $\SU(1,1)$. The $\SU(1,1)$ \6 symbol satisfies the
Lorentzian version of the asymptotics \Ref{asymp} \cite{asympt}, thus we expect to
be able to perform the same analysis pursued in this paper for the Riemannian case.
It is also worth mentioning that in the non--equilateral Lorentzian case, there are no
triangle inequalities to fulfill, thus the analysis is richer.

In light of these results, we think that this approach is very promising: it allows
computation of the leading order as well as the corrections of (a component of) the graviton
propagator, and, in a more general perspective, of semiclassical observables. When
applying this program to 4d, it will be interesting to compare and possibly match the
spinfoam corrections to the loop corrections computed in the standard perturbative
expansion (see e.g. \cite{Donoghue}). Let us nevertheless point out that the NNLO in our
3d model already contains an impressive number of terms and that the technical
difficulties to organize and compute further spinfoam corrections will likely be
comparable to the ones encountered when computing and summing the Feynman diagrams at
two--loop and higher orders.

Important issues remain open. The next step, to our opinion, is to be able to compute
all the components of the propagator, and recover the full tensorial structure.
This is crucial in 3d, in order to check satisfactory that the theory does not have local degrees of
freedom. The other fundamental question is the definition of a suitable coarse--graining procedure.
This is needed to understand how the single tetrahedron contribution is related to
the true large scale limit.
Both these issues require the extension of this formalism to deal with many tetrahedra.
We sketched how this can be done in Section II, and we leave this open for future work.
However, it is important to realize that increasing the number of tetrahedra makes the
numerical evaluations much more difficult.

\section*{Acknowledgements}

We would like to thank Carlo Rovelli for suggestions and encouragements.
We are grateful to Dan Christensen for useful discussions.  JLW was
supported by an International Research Fellowship of the NSF through
grant OISE--0401966, and is also grateful for the hospitality of the
Department of Mathematics at the University of Western Ontario.

\appendix

\section{Some elementary geometry}

In this appendix we report the geometric formulas used in this paper.
We begin by the derivatives of the Regge action $S_{\rm R}$, namely the matrix
$M_{ab}:=\f{\p S_{\rm R}}{\p j_1^a \p j_2^b}{\Big|_{j_e=j_0}}$
introduced in \Ref{Regge1}.
Recall that $\ell_e=j_e+\f12$, and
\equ
S_{\rm R}=\sum_e \ell_e \theta_e(\ell_e).
\nequ
All we need, to evaluate $M_{ab}$, is the dependence of the dihedral angles $\theta_e$ on
the edge lengths. To obtain this dependence, we start with the following well known
expression of the dihedral angles,
\equ\label{dih}
\sin\theta_e=\f32\f{\ell_e V}{A_1A_2},
\nequ
where $A_1$ and $A_2$ are the areas of the two triangles sharing the edge $e$.
All the quantities appearing above can be expressed in terms of the (squared) edge
lengths, using the Cayley matrix,
\begin{equation}
\label{cayley}
C_{(n)} =  \left(
\begin{array}{ccccc}
0        & 1         & 1             &   \ldots  &  1             \\
1        & 0         & \ell_1^2      &   \ldots  & \ell_n^2       \\
1        & \ell_1^2  & 0             &   \ldots  & \ell_{2n-1}^2  \\
\ldots   &   \ldots  &   \ldots      &   \ldots  & \ldots    \\
1        & \ell_n^2  & \ell_{2n-1}^2 &   \ldots  & 0
\end{array}
\right).
\end{equation}
Indeed, we have
\equ\label{nvolumeB}
V_{(n)}^2=\frac{(-1)^{n+1}}{2^n(n!)^2}\det C_{(n)}.
\nequ
Using this formula for $n=3$ to get the volume, and for $n=2$ to get the areas entering
\Ref{dih}, we can compute all the derivatives of the Regge action. For the equilateral
tetrahedron at hand, with $\ell_0=j_0+\f12$ and $\cos\vartheta=-\f13$, we have the
following values:
\equ
M_{ab}=\left( \begin{array}{ccccccc}
6\,\vartheta\, \ell_0 &  \vartheta & -\f{\sqrt 2}{3 \ell_0} & -\f{13}{9\sqrt2 \ell_0^2}
& -\f{209}{54\sqrt2 \ell_0^3} & -\f{1715}{108\sqrt2 \ell_0^4} & \ldots \\
\vartheta &  -\f{\sqrt 2}{\ell_0} & -\f1{\sqrt2 \ell_0^2} & -\f{5}{2\sqrt2 \ell_0^3}
& -\f{33}{4\sqrt2 \ell_0^4} & \\
-\f{\sqrt 2}{3 \ell_0} & -\f1{\sqrt2 \ell_0^2} & -\f{3}{2\sqrt2 \ell_0^3} &
-\f{21}{4\sqrt2 \ell_0^4} & & \\
-\f{13}{9\sqrt2 \ell_0^2} & -\f{5}{2\sqrt2 \ell_0^3} & -\f{21}{4\sqrt2 \ell_0^4} & & & \\
-\f{209}{54\sqrt2 \ell_0^3} & -\f{33}{4\sqrt2 \ell_0^4} & & & & \\
-\f{1715}{108\sqrt2 \ell_0^4} & & & & & & \\
\ldots &&&&&&\ldots\\
\end{array}\right).
\nequ
These are all the necessary quantities to compute the NLO, as in section \ref{SectionNLO},
and the NNLO, as in section \ref{SectionNNLO}.

In Section \ref{sectionrett}, we generalized the geometric setting and considered an
isosceles tetrahedron. In this case, we have two different dihedral
angles, corresponding to the space edges $\ell_0$, and the bulk edges $\ell_t$:
\equ
\cos{\theta_0}=-\f{{4 \ell_t^2-3\ell_0^2}}{4 \ell_t^2-\ell_0^2},\qquad
\sin{\theta_0} = 2\sqrt{2} \ell_0 \f{\sqrt{2 \ell_t^2 - \ell_0^2}}{4 \ell_t^2-\ell_0^2},
\nequ
and
\equ
\cos{\theta_t}=- \f{\ell_0^2}{4 \ell_t^2-\ell_0^2},
\qquad
\sin{\theta_t} = 2\sqrt{2} \ell_t \f{\sqrt{2 \ell_t^2 - \ell_0^2}}{4 \ell_t^2-\ell_0^2}.
\nequ
Using these, and proceeding as above, the matrix of derivatives of the Regge action is
\equ
\f{\p S_{\rm R}}{\p j_1^a \p j_2^b}{\Big|_{j_0, j_t}}=\left( \begin{array}{cccc}
4\,\theta_t\, \ell_t +2\theta_0 \ell_0 &  \theta_0 &
-\f{\ell_0^2}{4\ell_t^2-\ell_0^2}\f{\sqrt 2}{\sqrt{2\ell_t^2-\ell_0^2}} & \ldots \\
\theta_0 & -\f{\sqrt{2}}{\sqrt{2\ell_t^2-\ell_0^2}} & & \\
-\f{\ell_0^2}{4\ell_t^2-\ell_0^2}\f{\sqrt 2}{\sqrt{2\ell_t^2-\ell_0^2}} & & \ldots& \\
\ldots &&&\ldots\\
\end{array}\right).
\nequ
These are the only relevant terms for computing the LO \Ref{LOrett}.

\section{Bessel functions}

The modified Bessel functions of the first kind are defined by the following integral,
\equ\label{b1}
I_j(z)=\f1\pi\int_0^\pi d\phi \, e^{z \cos\phi} \cos(j \phi).
\nequ
Using this definition, it is straightforward to verify \Ref{norm} and \Ref{bessel}.
For the calculations of this paper, we have $z=1/\alpha$ are interested in the
$\alpha\mapsto 0$ limit, i.e. $z\mapsto\infty$. In this limit, we can evaluate \Ref{b1} by a saddle point
approximation. Firstly, we write the cosine in exponential form, so that
\Ref{b1} becomes a sum of two exponentials, with arguments $z\cos\phi\pm ij\phi$. Consider
the positive exponential first, $z\cos\phi+ij\phi$. The stationary point is given by
solving the equation $\sin \phi_0 = i \f jz$. To do so, we define $\phi_0=i\psi_0$, and
use $\sin(i\psi_0)= i \sinh(\psi_0)$, so that $\psi_0$ satisfies $\sinh\psi_0=\f jz\simeq
\psi_0$ in the $z\mapsto\infty$ limit. Consequently, $\cos\phi_0=\sqrt{1+(\f jz)^2}\simeq
1 +\f12 (\f jz)^2$. Expanding the exponent of \Ref{b1} around $\phi_0$ we thus obtain
\equ
z\left[\cos\phi_0 + i \f jz \phi_0 -\f12\cos\phi_0(\phi-\phi_0)^2\right]
\simeq z\left[1-\f12(\f jz)^2-\f12(\phi-i \f jz)^2\right].
\nequ
It is easy to check that the negative exponential gives the same result;
using it in \Ref{b1} we have
\equ\label{b2}
I_j(z)\simeq 2 e^{z} \,e^{-\f{j^2}{2z}} \, \f1{2\pi}\int_{-\infty}^\infty d\phi\, e^{-\f z2(\phi-i \f jz)^2}
= \sqrt{\f2{\pi z}}\,e^{z} \,e^{-\f{j^2}{2z}}.
\nequ
Consequently, in this limit $I_j(z)\gg I_{j+1}(z)$.

From these results, it is straightforward to derive \Ref{BesselGauss}.


\end{document}